\documentclass[12pt]{article}
\usepackage{latexsym}

\newcommand{\startappendix}{\appendix
  \renewcommand{\theequation}{\Alph{section}.\arabic{equation}}}
\renewcommand{\theequation}{\arabic{section}.\arabic{equation}}
\newcommand{\newsection}[1]{\section{#1}\setcounter{equation}{0}} 
\newcommand{\beq}{\begin{eqnarray}}
\newcommand{\eeq}{\end{eqnarray}}
\newcommand{\beqnn}{\begin{eqnarray*}}
\newcommand{\eeqnn}{\end{eqnarray*}}
\newtheorem{theorem}{Theorem}
\newtheorem{prop}{Proposition}

\newcommand{\proof}{\medskip\par\noindent {\it Proof.\/}\quad}
\newcommand{\qed}{$\Box$ \bigskip\par}

\newcommand{\rd}{\partial}

\newcommand{\Tr}{\mathop{\mathrm{Tr}}}

\newcommand{\tp}[1]{\;{}^{\mathrm{t}}#1}

\newcommand{\ZZ}{\mathbf{Z}}
\newcommand{\Atilde}{\tilde{A}}
\newcommand{\atilde}{\tilde{a}}

\newcommand{\calL}{\mathcal{L}}
\newcommand{\calM}{\mathcal{M}}
\begin{document}

%%%%%%%%%%%%%%%%
%% title page %% 
%%%%%%%%%%%%%%%%

\title{Spectral curve, Darboux coordinates and 
Hamiltonian structure of periodic dressing chains}
\author{Kanehisa Takasaki
{\normalsize Department of Fundamental Sciences}\\
{\normalsize Faculty of Integrated Human Studies, Kyoto University}\\
{\normalsize Yoshida, Sakyo-ku, Kyoto 606-8501, Japan}\\
{\normalsize E-mail: takasaki@math.h.kyoto-u.ac.jp}}
\date{}
\maketitle

\begin{abstract}
A chain of one-dimensional Schr\"odinger operators connected by 
successive Darboux transformations is called the ``Darboux chain'' 
or ``dressing chain''.   The periodic dressing chain with period 
$N$ has a control parameter $\alpha$.  If $\alpha \not= 0$, the 
$N$-periodic dressing chain may be thought of as a generalization 
of the fourth or fifth (depending on the parity of $N$) Painlev\'e 
equations .  The $N$-periodic dressing chain has two different 
Lax representations due to Adler and to Noumi and Yamada.  Adler's 
$2 \times 2$ Lax pair can be used to construct a transition matrix 
around the periodic lattice.  One can thereby define an associated 
``spectral curve'' and a set of Darboux coordinates called ``spectral 
Darboux coordinates''.  The equations of motion of the dressing chain 
can be converted to a Hamiltonian system in these Darboux coordinates.  
The symplectic structure of this Hamiltonian formalism turns out to 
be consistent with a Poisson structure previously studied by Veselov, 
Shabat, Noumi and Yamada.  
\end{abstract}

\vfill
\begin{flushleft}
arXiv:nlin.SI/0206049 
\end{flushleft}
\newpage

%%%%%%%%%%%%%%%
%% main text %% 
%%%%%%%%%%%%%%%

\newsection{Introduction}

A chain of one-dimensional Schr\"odinger operators 
$L_n$, $n \in \ZZ$, connected by successive Darboux 
transformations is called a ``Darboux chain'' or 
``dressing chain'' \cite{bib:Sh-Ya,bib:Shabat}. 
This is a kind of nonlinear lattice.  A periodic 
dressing chain has a control parameter $\alpha$.  
The nature of this system changes drastically 
whether $\alpha$ vanishes or not.  If $\alpha = 0$, 
the chain consists of finite-band operators 
\cite{bib:Novikov,bib:Lax,bib:Mc-vM};  
the structure of those operators can be 
described by algebro-geometric methods 
\cite{bib:DMN,bib:Krichever,bib:Dubrovin}.  
If $\alpha \not= 0$, the system is equivalent to 
the fourth and fifth Painlev\'e equations and their 
higher order analogues \cite{bib:Ve-Sh,bib:Adler1}; 
the generic solutions are expected to be transcendental 
and beyond algebro-geometric methods.   Thus the notion 
of periodic dressing chains invites us to an interesting 
world in which algebro-geometric functions and Painlev\'e 
transcendents coexist and are connected by a continuous 
parameter.  

Periodic dressing chains have been studied from 
the point of view of Hamiltonian or Poisson structures 
as well.  Veselov and Shabat \cite{bib:Ve-Sh} developed 
a bi-Hamiltonian formalism of periodic dressing chains. 
Noumi and Yamada \cite{bib:No-Ya1} took up substantially 
the same issue in the framework of their ``higher order 
Painlev\'e equations of $A^{(1)}_\ell$ type'' (referred to 
as {\it the Noumi-Yamada system} in the following).  
Although discovered on a quite different ground, 
the Noumi-Yamada system of $A^{(1)}_\ell$ type is actually 
equivalent to the periodic dressing chain with period 
$N = \ell + 1$.  Noumi and Yamada define a Poisson structure 
on the phase space of this system, which coincides with 
one of the Poisson structures in the bi-Hamiltonian structure 
of Veselov and Shabat.  

On the other hand, a conceptually different Hamiltonian 
formalism has been developed by Okamoto \cite{bib:Okamoto} 
for the Painlev\'e equations and Garnier's generalizations 
\cite{bib:Garnier}.  This Hamiltonian formalism is 
based on the notion of isomonodromic deformations of 
an ordinary differential equation.  Harnad and Wisse 
\cite{bib:Harnad,bib:Ha-Wi} pointed out that the notion 
of ``spectral Darboux coordinates'' \cite{bib:AHH} 
lies in the heart of this kind of Hamiltonian formalism 
of isomonodromic deformations.  This reveals a close 
relationship to ``separation of variables'' of 
integrable systems \cite{bib:Moser1,bib:Sklyanin}.  
Since periodic dressing chains include the fourth 
and fifth Painlev\'e equations, one will naturally 
ask whether periodic dressing chains have a similar 
Hamiltonian formalism.  This is the main problem that 
we address in this paper.  

For this purpose, we need an isomonodromic Lax formalism 
of periodic dressing chains.  Two options can be found 
in the literature.  One is a Lax pair of $2 \times 2$ 
matrices used by Veselov and Shabat \cite{bib:Ve-Sh} 
for the case of $\alpha = 0$ and modified by Adler 
\cite{bib:Adler2} to fit into the case with an arbitrary 
value of $\alpha$.  Strictly speaking, Adler's Lax pair 
is related to deformations of a {\it difference} (rather 
than differential) equation.  Therefore this is {\it not} 
an isomonodromic system in the usual sense.  
Another option is a Lax pair of $N \times N$ matrices 
presented by Noumi and Yamada\cite{bib:No-Ya2}.  
This is an isomonodromic system in the usual sense. 
These two Lax pairs turn out to be equivalent 
(or {\it dual\/}) and connected with each other 
by a Mellin transformation.  

We use Adler's Lax pair and construct a $2 \times 2$ 
transition matrix $T(\lambda)$ around the periodic chain.  
This matrix satisfies a Lax equation, which inherits 
the unusual nature of Adler's Lax pair.  Nevertheless, 
we can construct a hyperelliptic ``spectral curve'' 
$\Gamma$ thereof, and convert the Lax equation to 
a dynamical system of a finite number of moving points 
$(\lambda_j,z_j) \in \Gamma$, $j = 1,\ldots,g$.   
Another complicated aspect of this system is that 
the spectral curve itself is dynamical in the case 
where $\alpha \not= 0$, but this is rather a universal 
characteristic of isomonodromic systems.  The coordinates 
$\lambda_j,z_j$ of the moving points are nothing but 
``spectral Darboux coordinates'' mentioned above.  
The equations of motion of the moving points 
becomes a Hamiltonian system (which we call, 
for convenience, {\it the spectral Hamiltonian system}) 
in these new coordinates, and conversely, the Lax equation
can be reconstructed from this Hamiltonian system.  

It is not obvious from the construction that 
the passage from the periodic dressing chain to 
the spectral Hamiltonian system is invertible.  
The problem is to reconstruct the phase space 
coordinates of the periodic dressing chain 
or, equivalently, of the Noumi-Yamada system 
from the spectral Darboux coordinates.  
We solve this {\it inverse problem} and find  
the existence of a locally invertible map 
that connects the spectral Hamiltonian system with 
the Noumi-Yamada system directly.  As a byproduct, 
we can show that this map is a Poisson map connecting 
two apparently different odd-dimensional Poisson 
structures. 

This paper is organized as follows.  
Sections 2 and 3 review the equations of 
motion of dressing chains, the Noumi-Yamada systems, 
and Lax pairs of these systems. Section 4 deals with 
the transition matrix and the associated spectral curve.  
Sections 5 and 6 present the main results of this paper.  
Section 5 is concerned with the Hamiltonian system 
in spectral Darboux coordinates.  Section 6 is devoted 
to the inverse problem and Poisson structures. 
Section 7 is for concluding remarks.  Some technical 
details are collected in Appendices.

\newsection{Dressing chain}

\subsection{Definition}

A dressing chain consists of one-dimensional 
Schr\"odinger operators of the factorized form 
\beq
  L_n = (\rd_x - v_n)(\rd_x + v_n),  \; 
  \rd_x = \rd/\rd x, 
\eeq
that are linked with the neighbors by 
the Darboux transformations 
\beq
  (\rd_x - v_{n+1})(\rd_x + v_{n+1}) 
  = (\rd_x + v_n)(\rd_x - v_n) + \alpha_n 
\eeq
with parameters $\alpha_n$.  
This is a kind of nonlinear lattice governed 
by the differential equations 
\beq
  \dot{v}_n + \dot{v}_{n+1} = v_{n+1}^2 - v_n^2 + \alpha_n, 
  \label{eq:dchain}
\eeq
where the dot stands for the $x$-derivative: 
\beqnn
  \dot{v}_n = \rd_x v_n. 
\eeqnn

We are mostly interested in periodic dressing 
chains satisfying  the periodicity condition 
\beq
  v_{n+N} = v_n, \; \alpha_{n+N} = \alpha_n. 
\eeq
Since the case of $N = 2$ is no very interesting, 
the subsequent consideration is focussed on 
the case where $N \ge 3$.  The auxiliary variable 
\beq
  v = \sum_{n=1}^N v_n 
    = \frac{1}{2}\sum_{n=1}^N (v_n + v_{n+1}) 
\eeq
satisfies the very simple equation of motion 
\beq
  \dot{v} = \frac{\alpha}{2}, 
  \label{eq:dot-v}
\eeq
where $\alpha$ is the constant defined by 
\beq
  \alpha = \sum_{n=1}^N \alpha_n. 
\eeq
Thus $v$ is a linear function of 
the form $(\alpha/2)x + \mbox{constant}$.  
This means that $v$ is not a dynamical variable 
in the usual sense.  

The parameter $\alpha$ is a control parameter 
of the periodic dressing chain.   
To see how the situation depends on this parameter, 
we note that the $N$-th order operator 
\beq
  M_n = (\rd_x + v_{n+N-1}) \cdots (\rd_x + v_n) 
\eeq
satisfies the equation 
\beq
  [L_n,M_n] = \alpha M_n. 
  \label{eq:[Ln,Mn]}
\eeq
If $\alpha = 0$, $L_n$ and $M_n$ are commuting 
differential operators in the sense of 
Burchnall and Chaundy \cite{bib:Bu-Ch}.  
In particular, the case of odd $N$'s falls 
into the classical theory of finite-band operators.  
The case of even $N$'s is more involved, because 
the orders of the operators are not co-prime.  
If $\alpha \not= 0$, the operator equation 
resembles the so called ``string equations'' 
of non-critical string theory and 
two-dimensional quantum gravity \cite{bib:Moore}.  
The usual string equations take the form 
\beq
  [L,M] = \alpha 
\eeq
(scaling limit of Hermitian matrix models) or 
\beq
  [L,M] = \alpha L 
\eeq
(scaling limit of unitary matrix models). 
In particular, if $L$ is a second order 
Schr\"odinger operator $L = \rd_x^2 - u$, 
the string equations of the first type give 
the first Painlev\'e equation and its higher order 
analogues.  The string equation of the second type 
are related to the second Painlev\'e equation.  
Similarly, the periodic dressing chains for 
$N = 3$ and $N = 4$ are equivalent to the fourth 
and fifth Painlev\'e equations, as pointed out by 
Veselov and Shabat \cite{bib:Ve-Sh} and 
Adler \cite{bib:Adler1}.

\subsection{Relation to Noumi-Yamada systems}

The periodic chains have another expression 
in terms of the new dependent variables 
\beq
  f_n = v_{n+1} + v_n. 
  \label{eq:fn-vn}
\eeq
As we shall see below, this gives a change of variables 
to the Noumi-Yamada system \cite{bib:No-Ya1}.

\subsubsection{If $N$ is odd}

If $N = 2g + 1$, (\ref{eq:fn-vn}) can be solved for $v_n$ as 
\beq
  v_n = \frac{1}{2}(f_n - f_{n+1} + \cdots + f_{n+2g}). 
\eeq
In this case, $v_n$'s and $f_n$'s are thus connected 
by an invertible linear map.  Rewriting (\ref{eq:dchain}) as 
\beqnn
  \dot{f}_n = f_n(v_{n+1} - v_n) + \alpha_n 
  \label{eq:dot-fn}
\eeqnn
and inserting this expression of $v_n$ therein, 
one obtains the closed differential equations 
\beq
  \dot{f}_n 
  = f_n \left(\sum_{k=1}^g f_{n+2k-1} 
            - \sum_{k=1}^g f_{n+2k}\right) 
    + \alpha_n 
  \label{eq:NY-Nodd}
\eeq
for $f_n$'s.  This is exactly the Noumi-Yamada system 
of $A^{(1)}_{2g}$ type.

\subsubsection{If $N$ is even}

If $N = 2g + 2$, (\ref{eq:fn-vn}) cannot be solved 
for $v_n$'s uniquely.  This ambiguity, however, 
turns out to be removed by the existence of 
an extra constraint. This constraint is 
a consequence of the obvious identity 
\beq
  \sum_{n=1}^{2g+2}(-1)^n(\dot{v}_n + \dot{v}_{n+1}) = 0 
  \label{eq:dot-vn-identity}
\eeq
and the equations of motion of the dressing chain.  
Because of this extra constraint, $v_n$'s and 
$f_n$'s are connected by a birational map; 
see Appendix B for details.  Thus one can, 
in principle, rewrite the equations of motion 
of the dressing chain into a system of 
differential equations for $f_n$'s.  
We can, however, circumvent messy calculations 
that will be inevitable if we do the change of 
variables naively, as follows.  

The clue is to calculate the derivative of $vf_n$ 
rather than $f_n$ itself.  Since 
\beqnn
  v = \sum_{k=1}^{g+1}f_{n+2k-1}, 
\eeqnn  
the derivative of $vf_n$ can be written as 
\beqnn
  (vf_n)\dot{\;}
  = \sum_{k=1}^{g+1}(\dot{f}_{k+2k-1}f_n + f_{n+2k-1}\dot{f}_n). 
\eeqnn
We can now use (\ref{eq:dchain}) to rewrite 
each term of the sum on the right hand side as 
\beqnn
  \lefteqn{\dot{f}_{n+2k-1}f_n + f_{n+2k-1}\dot{f}_n} 
  \nonumber \\
  &=& (f_{n+2k-1}(v_{n+2k} - v_{n+2k-1}) + \alpha_{n+2k-1})f_n  
      + f_{n+2k-1}(f_n(v_{n+1} - v_n) + \alpha_n) 
  \nonumber \\
  &=& f_{n+2k-1}f_n(f_{n+2k-1} - 2f_{n+2k-2} + \cdots 
        + 2f_{n+1} - f_n) 
  \nonumber \\
  && \mbox{} 
      + \alpha_{n+2k-1}f_n + \alpha_n f_{n+2k-1}. 
\eeqnn
Suming over $n$ and doing some algebra, we eventually 
obtain the differential equations 
\beq
  (vf_n)\dot{\ }
  &=& f_n \left(\sum_{1\le j\le k\le g}f_{n+2j-1}f_{n+2k} 
              - \sum_{1\le j\le j\le g}f_{n+2j}f_{n+2k+1}\right)
  \nonumber\\
  && \mbox{}
    + \sum_{k=1}^{g+1} \alpha_{n+2k-1} f_n 
    + \alpha_n v  
\eeq
or, equivalently, 
\beq
  v\dot{f}_n 
  &=& f_n \left(\sum_{1\le j\le k\le g}f_{n+2j-1}f_{n+2k} 
              - \sum_{1\le j\le k\le g}f_{n+2j}f_{n+2k+1}\right)
  \nonumber\\
  && \mbox{}
    + \left(\sum_{k=1}^{g+1} \alpha_{n+2k-1} 
          - \frac{\alpha}{2}\right) f_n 
    + \alpha_n v. 
  \label{eq:NY-Neven}
\eeq
If $v$ is normalized as $v = \alpha x / 2$, 
this is nothing but the Noumi-Yamada system of type 
$A^{(1)}_{2g+1}$.  Note that the variables $f_n$, 
being defined by (\ref{eq:fn-vn}), also satisfy 
the constraint 
\beq
  \sum_{k=1}^{g+1} f_{2k-1} = \sum_{k=1}^{g+1} f_{2k}. 
  \label{eq:fn-constraint}
\eeq

\newsection{Lax pairs}

The dressing chains have two different Lax pairs, 
one due to Veselov and Shabat \cite{bib:Ve-Sh} 
and Adler \cite{bib:Adler2}, and the other 
due to Noumi and Yamada \cite{bib:No-Ya2}.  
We point out that these two Lax pairs are 
connected by a Mellin transformation.

\subsection{$2 \times 2$ Lax pair}

The Lax pair of Veselov and Shabat is valid only 
for the case of $\alpha = 0$.  Adler proposed 
a slightly modified Lax pair in which this assumption 
is removed.  

Let $u_n$ denote the function 
\beq
  u_n = v_n^2 - \dot{v}_n.  
\eeq
This is the  ``potential'' of 
the Schr\"odinger perator $L_n$: 
\beq
  L_n = \rd_x^2 - u_n. 
\eeq
Adler's Lax pair consists of the $2 \times 2$ matrices 
\beq
  V_n(\lambda) 
  = \left(\begin{array}{cc}
    v_n & 1 \\
    \lambda + v_n^2 & v_n 
    \end{array}\right)
\eeq
and 
\beq
  U_n(\lambda) 
  = \left(\begin{array}{cc} 
    0  & 1 \\
    \lambda + u_n & 0 
    \end{array}\right), 
\eeq
which satisfy the differential equation 
\beq
  \dot{V}_n(\lambda) 
  = U_{n+1}(\lambda + \alpha_n) V_n(\lambda)  
  - V_n(\lambda) U_n(\lambda).  
  \label{eq:dot-Vn} 
\eeq
This ``Lax'' equation resembles the Lax equation 
of many integrable nonlinear lattices; an unusual 
feature is that it contains a shift of the spectral 
parameter. 

An associated auxiliary linear system takes the form 
\beq
  \dot{\Phi}_n(\lambda) = U_n(\lambda)\Phi_n(\lambda), 
  \label{eq:dot-Phin} \\ 
  \Phi_{n+1}(\lambda + \alpha_n) 
  + V_n(\lambda)\Phi_n(\lambda) = 0. 
  \label{eq:diff-Phin} 
\eeq
The Lax equation (\ref{eq:dot-Vn}) can be 
reconstructed from these linear equations 
as the consistency (integrability) condition. 
One can rewrite this matrix linear system to 
a scalar form.  Firstly, (\ref{eq:dot-Phin}) 
implies that $\Phi_n(\lambda)$ can be expressed as 
\beqnn
  \Phi_n(\lambda) 
  = \left(\begin{array}{cc} 
    \phi_n(\lambda) \\
    \dot{\phi}_n(\lambda) 
    \end{array}\right), 
\eeqnn
and that the scalar function $\phi_n(\lambda)$ 
satisfies the Schr\"odinger equation 
\beq
  L_n \phi_n(\lambda) 
  = \ddot{\phi}_n(\lambda) - u_n \phi_n(\lambda) 
  = \lambda \phi_n(\lambda). 
  \label{eq:ddot-phin}
\eeq
Secondly, (\ref{eq:diff-Phin}) yields 
the the scalar difference-differential equation 
\beq
  \phi_{n+1}(\lambda + \alpha_n) 
  + \dot{\phi}_n(\lambda) 
  + v_n\phi_n(\lambda) = 0. 
  \label{eq:diff-dot-phin}
\eeq

\subsection{Another Lax pair}

Another Lax pair emerges from the auxiliary linear system 
\beq
  \alpha z\rd_z \psi_n(z) + \psi_{n+2}(z) 
  + f_n\psi_{n+1}(z) + e_n\psi_n(z) = 0, 
  \label{eq:zdz-psin} \\ 
  \dot{\psi}_n(z) + \psi_{n+1}(z) + v_n\psi_n(z) = 0. 
  \label{eq:dot-psin} 
\eeq
Since (\ref{eq:dot-psin}) implies that 
\beqnn
  \lefteqn{\psi_{n+2}(z) + f_n\psi_{n+1}(z) + e_n\psi_n(z)}
     \nonumber \\
  &=& (\rd_x + v_{n+1})(\rd_x + v_n)\psi_n(z) 
     - (v_n + v_{n+1})(\rd_x + v_n)\psi_n + e_n\psi_n(z) 
     \nonumber \\
  &=& (L_n + e_n)\psi_n(z), 
\eeqnn
one can rewrite (\ref{eq:zdz-psin}) as 
\beq
  \alpha z\rd_z \psi_n(z) + (L_n + e_n)\psi_n(z) = 0. 
  \label{eq:zdz-psin2}
\eeq

The consistency condition of these linear equations 
yields the three equations 
\beq
  f_{n+1} - f_n = v_{n+2} - v_n, 
  \label{eq:psin-compat1} \\
  \dot{f}_n = f_n(v_{n+1} - v_n) + e_n - e_{n+1}, 
  \label{eq:psin-compat2}\\
  \dot{e}_n = 0. 
\eeq

These equations are retained under 
{\it gauge transformations} 
\beq
  \psi_n \to \psi_n e^\gamma, \quad 
  v_n \to v_n - \dot{\gamma}, \quad 
  f_n \to f_n, \quad 
  e_n \to e_n. 
\eeq
The equations of motion (\ref{eq:dchain}) 
of the infinite dressing chain can be 
recovered by a special ``gauge fixing'' in which 
(\ref{eq:fn-vn}) is satisfied.  Let us rewrite 
the first equation (\ref{eq:psin-compat1}) of 
the consistency condition as 
\beqnn
  f_{n+1} - v_{n+2} - v_{n+1} = f_n - v_{n+1} - v_n. 
\eeqnn
Both hand sides of this equation is thus independent 
of $n$, thereby become a function $h = h(x)$ of 
$x$ only.  This function transforms as 
$h \to h + 2\dot{\gamma}$ under the gauge transformation, 
and can be eliminated (or ``gauged away'') by 
a suitable choice of the function $\gamma$.  
One can thus achieve a special gauge that satisfies 
(\ref{eq:fn-vn}).  In this gauge, 
the second equation (\ref{eq:psin-compat2}) of 
the consistency condition becomes the equation 
(\ref{eq:dchain}) of the dressing chain 
upon identifying the parameters $\alpha_n$ as 
\beq
  \alpha_n = e_n - e_{n+1}. 
\eeq
Note that $\alpha$ is so far a free parameter; 
the relation to $\alpha_n$'s emerges after 
imposing the $N$-periodicity condition 
(see below.)  

This auxiliary linear system reduces to 
that of Noumi and Yamada \cite{bib:No-Ya2} 
for higher order Painlev\'e equations 
by imposing the periodicity 
conditions 
\beq
  v_{n+N} = v_n, \; f_{n+N} = f_n, \; e_{n+N} = e_n - \alpha. 
\eeq
The last condition implies the relation 
\beqnn
  \alpha = \sum_{n=1}^N (e_n - e_{n+1}) 
         = \sum_{n=1}^N \alpha_n 
\eeqnn
as expected.  We further assume the quasi-periodicity 
\beq
  \psi_{n+N}(z) = z \psi_n(z) 
  \label{eq:psin-bc}
\eeq
of $\psi_n(z)$, $z$ being interpreted 
to be the Bloch-Floquet multiplier.  
The auxiliary linear system of the 
infinite chain thereby turns into 
the $N \times N$ matrix equations 
\beq
  \alpha z\rd_z \Psi(z) + \calL(z)\Psi(z) = 0, \; 
  \dot{\Psi}(z) + \calM(z) \Psi(z) = 0 
\eeq
for the column vector 
\beqnn
  \Psi(z) 
  = \tp{\Bigl(\psi_1(z) &\cdots&\psi_N(z)\Bigr)}, 
\eeqnn
where $\calL(z)$ and $\calM(z)$ denotes 
the $N \times N$ matrices 
\beq
  \calL(z) &=& 
    \left(\begin{array}{ccccc}
    e_1 & f_1 & 1 &   &  \\
        &\ddots&\ddots&\ddots& \\
        &      &\ddots&\ddots&1 \\
    z   &      &      &e_{N-1}&f_{N-1}\\
    f_Nz & z   &      &      &e_N 
    \end{array}\right), 
  \\ 
  \calM(z) &=& 
    \left(\begin{array}{cccc}
    v_1 & 1 &   &   \\
        &\ddots&\ddots& \\
        &      &\ddots&1 \\
    z   &      &      &v_N 
    \end{array}\right). 
\eeq
This Lax pair may be thought of as 
a kind of self-similarity reduction of 
the Drinfeld-Sokolov systems \cite{bib:Dr-So}.

\subsection{Mellin transformation connecting two Lax pairs}

The two auxiliary linear systems turn out to 
be connected by the Mellin transformation 
\beq
  \psi_n(z) 
  = \int z^{-\lambda/\alpha}\phi_n(\lambda - e_n)d\lambda 
\eeq
provided that $\alpha \not= 0$.  (\ref{eq:dot-psin}) 
can be derived from (\ref{eq:diff-dot-phin}) as 
\beqnn
  \psi_{n+1}(z) 
  &=& \int z^{-\lambda/\alpha}
      \phi_{n+1}(\lambda - e_n + \alpha_n)d\lambda 
  \nonumber \\ 
  &=& - \int z^{-\lambda/\alpha}
        (\rd_x + v_n)\phi_n(\lambda - e_n)d\lambda 
  \nonumber \\
  &=& - (\rd_x + v_n)\psi_n(z). 
\eeqnn
Similarly, (\ref{eq:zdz-psin2}) is a consequence of 
the identity 
\beqnn
  \alpha z\rd_z \psi_n(z) 
  = - \int z^{-\lambda/\alpha}\lambda\phi_n(\lambda - e_n)d\lambda 
\eeqnn
and the equation 
\beqnn
  \lambda\phi_n(\lambda - e_n) 
  &=& (\lambda - e_n)\phi_n(\lambda - e_n) 
    + e_n\phi_n(\lambda - e_n) 
  \nonumber \\
  &=& (L_n + e_n)\phi_n(\lambda - e_n) 
\eeqnn
that can be derived from (\ref{eq:ddot-phin}).  

Note that the Mellin transformation is a {\it heuristic} 
relation rather than a rigorous one.  For instance, 
we have not specified the meaning of the integral.  
Nevertheless, this correspondence is very useful 
for understanding the origin of the two different 
Lax pairs.  Also note that the Mellin transformation 
can be converted to a Laplace transformation by 
changing the variable from $z$ to $\zeta = \log z$.

\newsection{Transition matrix and spectral curve}

We construct the transition matrix $T(\lambda)$ of 
periodic dressing chains.  This matrix satisfies 
a Lax equation, which is not isospectral if 
$\alpha \not= 0$.  The spectral curve $\Gamma$ 
is accordingly ``time-dependent'', i.e., 
deforms as $x$ varies.

\subsection{Transition matrix}

Following the usual prescription, we can consider 
an analogue of the transition matrices around 
the periodic lattice.  The transition matrix 
$T(\lambda)$ that connecting $\Phi_1$ with $\Phi_{N+1}$ 
is defined by the linear relation 
\beq
  \Phi_{N+1}(\lambda + \alpha) 
  = (-1)^N T(\lambda)\Phi_1(\lambda).  
\eeq
Note that $\lambda$ is also shifted 
as $\lambda \to \lambda + \alpha$.  
More explicitly, $T(\lambda)$ can be written as 
\beq
  T(\lambda) 
  = V_N(\lambda + \beta_{N-1}) \cdots 
    V_2(\lambda + \beta_1) V_1(\lambda), 
\eeq
where 
\beq
  \beta_n = \alpha_1 + \alpha_2 + \cdots + \alpha_n, \; 
  \beta_0 = 0, 
\eeq
and satisfies the Lax equation 
\beq
  \dot{T}(\lambda) 
  = U_1(\lambda + \alpha)T(\lambda) - T(\lambda)U_1(\lambda). 
  \label{eq:dot-T}
\eeq
Note that this Lax equation is {\it not} isospectral 
if $\alpha \not= 0$.  In terms of the matrix elements, 
\beqnn
  T(\lambda) 
  = \left(\begin{array}{cc}
    A(\lambda) & B(\lambda) \\
    C(\lambda) & D(\lambda) 
    \end{array}\right), 
\eeqnn
the Lax equation of $T(\lambda)$ reads 
\beq
  \dot{A}(\lambda) 
  &=& C(\lambda) - B(\lambda)(\lambda + u_1), 
  \label{eq:dot-A} \\
  \dot{B}(\lambda) 
  &=& D(\lambda) - A(\lambda), 
  \label{eq:dot-B} \\
  \dot{C}(\lambda) 
  &=& (\lambda + \alpha + u_1)A(\lambda) 
    - D(\lambda)(\lambda + u_1), 
  \label{eq:dot-C} \\
  \dot{D}(\lambda) 
  &=& (\lambda + \alpha + u_1)B(\lambda) - C(\lambda). 
  \label{eq:dot-D} 
\eeq

\subsection{Equation of spectral curve}

The affine part of the spectral curve $\Gamma$ 
is defined by the equation 
\beq
  \det(zI - T(\lambda)) = z^2 - P(\lambda)z + Q(\lambda) = 0 
\eeq
on the $(\lambda,z)$ plane. $P(\lambda)$ and $Q(\lambda)$ 
are the spectral invariants 
\beq
  P(\lambda) = \Tr T(\lambda), \; 
  Q(\lambda) = \det T(\lambda) 
\eeq
of $T(\lambda)$.  Since $T(\lambda)$ is a product of 
$2 \times 2$ matrices, $Q(\lambda)$ can be readily determined: 
\beq
  Q(\lambda) = (-1)^N \lambda \prod_{n=1}^{N-1}(\lambda + \beta_n). 
\eeq

To see what $P(\lambda)$ looks like, one has to specify 
the matrix elements of $T(\lambda)$ in more detail.   
The following can be confirmed rather easily by induction 
on $N$: 
\begin{enumerate}
\item If $N = 2g + 1$, the matrix elements of $T(\lambda)$ 
are polynomials of the form 
\beq
 \begin{array}{rcl}
  A(\lambda) &=& a_0\lambda^g + a_1\lambda^{g-1} + \cdots + a_g, \\
  B(\lambda) &=& b_0\lambda^g + b_1\lambda^{g-1} + \cdots + b_g, \\ 
  C(\lambda) &=& c_0\lambda^{g+1} + c_1\lambda^g + \cdots + c_{g+1}, \\ 
  D(\lambda) &=& d_0\lambda^g + d_1\lambda^{g-1} + \cdots + d_g, 
 \end{array}
\eeq
where 
\beq
  b_0 = c_0 = 1, \; a_0 = d_0 = v. 
\eeq
Consequently, $P(\lambda)$ is a polynomial of the form 
\beq
  P(\lambda) = I_0 \lambda^g + I_1\lambda^{g-1} + \cdots + I_g. 
\eeq
\item If $N = 2g + 2$, the matrix elements of $T(\lambda)$ 
are polynomials of the form 
\beq
 \begin{array}{rcl}
  A(\lambda) &=& \lambda^{g+1} + a_0\lambda^g + \cdots + a_g, \\
  B(\lambda) &=& b_0 \lambda^g + b_1\lambda^{g-1} + \cdots + b_g, \\
  C(\lambda) &=& c_0\lambda^{g+1} + c_1\lambda^g + \cdots + c_{g+1}, \\
  D(\lambda) &=& \lambda^{g+1} + d_0\lambda^g + d_1\lambda^{g-1} 
      + \cdots + d_g,
 \end{array}
\eeq
where 
\beq
  b_0 = c_0 = v. 
\eeq
Consequently, $P(\lambda)$ is a polynomial of the form 
\beq
  P(\lambda) 
  = 2\lambda^{g+1} + I_0 \lambda^g + I_1\lambda^{g-1} + \cdots + I_0. 
\eeq
\end{enumerate}

$\Gamma$ is a hyperelliptic curve of genus $g$.  
The defining equation can be transformed to 
the normal form 
\beq
  y^2 = P(\lambda)^2 - 4Q(\lambda) 
\eeq
by changing the variables as 
$z = (P(\lambda) + y)/2$. 
If $\alpha = 0$, the coefficients $I_n$ of $P(\lambda)$ 
are constants of motion; if $\alpha \not= 0$, 
they depend on $x$.  This is obvious from 
the equation 
\beq
  \dot{P}(\lambda) = \alpha B(\lambda), 
  \label{eq:dot-P}
\eeq
which can be derived from (\ref{eq:dot-A}) and 
(\ref{eq:dot-D}). 

A consequence of (\ref{eq:dot-P}) is the equation 
\beq
  \dot{I}_0 = \alpha b_0. 
\eeq
Since $b_0$ is a constant (for odd $N$'s) or 
a simple linear function of $x$ (for even $N$'s), 
this equation shows that $I_0$, too, is 
a simple (at most quadratic) function of $x$ 
even if $\alpha \not= 0$.  We shall present 
a more explicit expression of $I_0$ below.

One can now find an explicit expression of 
the coefficients $a_0$ and $d_0$ for $N = 2g + 2$.  
This is a consequence of (\ref{eq:dot-B}); 
this equation implies that 
\beq
  \dot{b}_0 = d_0 - a_0. 
\eeq
Since $d_0 + a_0 = I_0$ and 
$\dot{b}_0 = \dot{v} = \alpha/2$, 
this immediately leads to the expression 
\beq
  a_0 = \frac{1}{2}(I_0 - \dot{b}_0) 
      = \frac{1}{2}(I_0 - \frac{\alpha}{2}), \; 
  d_0 = \frac{1}{2}(I_0 + \dot{b}_0) 
      = \frac{1}{2}(I_0 + \frac{\alpha}{2}) 
\eeq
of $a_0$ and $d_0$ in terms of $I_0$.  

This expression implies that $a_0$ and $d_0$, 
like $b_0$, $c_0$ and $I_0$, are a simple 
(at most quadratic) function of $x$ 
(in particular, a constant if $\alpha = 0$).  
We shall interpret these quantities as 
Casimir functions of a Poisson structure.

\subsection{Structure of $P(\lambda)$ in more detail}

We here review an explicit formula of $P(\lambda)$ 
presented by Veselov and Shabat \cite{bib:Ve-Sh}.  

By definition, $P(\lambda)$ is given by the trace 
\beq
  P(\lambda) 
  = \Tr V_N(\lambda + \beta_{N-1}) \cdots 
    V_2(\lambda + \beta_1)V_1(\lambda). 
  \label{eq:P-as-trace}
\eeq
We now split each matrix 
$V_n(\lambda + \beta_{n-1})$ in the trace 
into two rank-one matrices: 
\beq
  V_n(\lambda + \beta_{n-1}) 
  = \left(\begin{array}{c} 
    1 \\
    v_n
    \end{array}\right)
    \left(\begin{array}{cc}
    v_n & 1 
    \end{array}\right) 
  + \left(\begin{array}{c} 
    0 \\
    \lambda +\beta_{n-1} 
    \end{array}\right)
    \left(\begin{array}{cc}
    1 & 0 
    \end{array}\right). 
  \label{eq:split-Vn}
\eeq
For convenience, let us call the two factors 
on the right hand side as {\it type I} and 
{\it type II} factors, respectively.  

The trace is now expanded into a sum of 
$2^N$ terms, each term being the trace of 
a product of rank-one factors as above.  
The trace of a product of such matrices 
factorizes in itself. For instance, 
if all the rank-one factors in the trace 
is of type I, 
\beqnn
  \Tr   
    \left(\begin{array}{c}
    1 \\
    v_N 
    \end{array}\right) 
    \left(\begin{array}{cc}
    v_N & 1 
    \end{array}\right) 
    \cdots 
    \left(\begin{array}{c}
    1 \\
    v_1 
    \end{array}\right)
    \left(\begin{array}{cc}
    v_1 & 1 
    \end{array}\right) 
  &=& (v_N + v_{N-1})\cdots(v_2 + v_1)(v_1 + v_N) 
  \nonumber \\
  &=& f_1 f_2 \cdots f_N. 
\eeqnn
The expansion of $P(\lambda)$ starts with this term, 
and proceeds to higher order terms including type II 
factors.  General terms obey the following rules. 
\begin{enumerate}
\item 
A higher order term in this expansion vanishes 
if it includes a pair of {\it neighboring} factors 
of type II.  This is due to the obvious identity 
\beqnn
    \left(\begin{array}{c} 
    0 \\
    \lambda +\beta_{n-1} 
    \end{array}\right)
    \left(\begin{array}{cc}
    1 & 0 
    \end{array}\right) 
    \left(\begin{array}{c} 
    0 \\
    \lambda +\beta_{n-2} 
    \end{array}\right)
    \left(\begin{array}{cc}
    1 & 0 
    \end{array}\right) 
  = \left(\begin{array}{cc} 
    0 & 0 \\
    0 & 0 
    \end{array}\right). 
\eeqnn
\item After factorization, 
non-vanishing traces take place 
in the following three forms: 
\beqnn
  \Tr\left(\begin{array}{cc}
     v_{n+1} & 1 
     \end{array}\right)
     \left(\begin{array}{c}
     1 \\
     v_n 
     \end{array}\right) 
  &=& f_n, 
  \\
  \Tr\left(\begin{array}{cc}
     v_{n+1} & 1 
     \end{array}\right)
     \left(\begin{array}{c}
     0 \\
     \lambda + \beta_{n-1}
     \end{array}\right) 
  &=& \lambda + \beta_{n-1}, 
  \\
  \Tr\left(\begin{array}{cc}
     1 & 0 
     \end{array}\right) 
     \left(\begin{array}{cc}
     1 \\
     v_n 
     \end{array}\right)
  &=& 1. 
\eeqnn
In particular, $v_n$ always appears along 
with $v_{n\pm 1}$ in the linear combinations 
$v_n + v_{n+1} = f_n$ and $v_n + v_{n-1} = f_{n-1}$. 
\end{enumerate}

This expansion can be reorganized into 
the beautiful formula 
\beq
  P(\lambda) 
  = \prod_{n=0}^{N-1} \left(
    1 + (\lambda + \beta_n)
        \frac{\rd^2}{\rd f_n \rd f_{n+1}}
    \right)(f_1 f_2 \cdots f_N) 
  \label{eq:P-fn}
\eeq
due to Veselov and Shabat \cite{bib:Ve-Sh}.  
Here it is understood that $f_0 = f_N$ and 
$\beta_0 = 0$.  One can read off some of 
the coefficients of $P(\lambda)$ explicitly.  
In particular, $I_0$ turns out to be a polynomial 
in $v$ with constant coefficients: 
\begin{enumerate}
\item If $N = 2g + 1$, 
\beq
  I_0 = f_1 + f_2 + \cdots + f_{2g+1} = 2v. 
\eeq
\item If $N = 2g + 2$, 
\beq
  I_0 = \sum_{j,k=1}^{g+1} f_{2j-1}f_{2k} 
        + \sum_{n=1}^{2g+1} \beta_n 
      = v^2 + \sum_{n=1}^{2g+1} \beta_n. 
\eeq
\end{enumerate}

\newsection{Spectral Darboux coordinates and Hamiltonian system} 

The Lax equation of the transition matrix can be converted 
to a dynamical system of a finite number of points on 
the spectral curve.  The coordinates of these points 
are ``spectral Darboux coordinates'' in the terminology 
of the Montreal group \cite{bib:AHH,bib:Harnad,bib:Ha-Wi}.  
The dynamical system turns out to take a Hamiltonian form 
in these Darboux coordinates.

\subsection{Spectrl Darboux coordinates} 

Let $\lambda_j$, $j = 1,\cdots,g$, denote the zeroes
of $B(\lambda)$: 
\beq
  B(\lambda) = b_0 \prod_{j=1}^g (\lambda - \lambda_j). 
\eeq
We consider the generic case where the zeroes $\lambda_j$ 
are mutually distinct, and derive a system of differential 
equations for these zeroes. 

Differentiating the identity $B(\lambda_j) = 0$ 
by $x$ yields the identity 
\beqnn
  \dot{B}(\lambda_j) + B'(\lambda_j)\dot{\lambda}_j = 0, 
\eeqnn
where $B'(\lambda)$ stands for the $\lambda$-derivative: 
\beqnn
  B'(\lambda) = \rd_\lambda B(\lambda). 
\eeqnn
Solving this identity for $\dot{\lambda}_j$ and 
recalling (\ref{eq:dot-B}), we have 
\beq
  \dot{\lambda}_j 
  = \frac{A(\lambda_j) - D(\lambda_j)}{B'(\lambda_j)}. 
  \label{eq:pre-Dub}
\eeq
In order to rewrite the numerator of this equation 
in a more familiar form, let us notice the identity 
\beqnn
  (A(\lambda) - D(\lambda))^2 
  = P(\lambda)^2 - 4A(\lambda)D(\lambda). 
\eeqnn
Upon setting $\lambda = \lambda_j$, this identity 
becomes 
\beqnn
  (A(\lambda_j) - D(\lambda_j))^2 
  = P(\lambda_j)^2 - 4Q(\lambda_j) 
\eeqnn
because 
\beq
  A(\lambda_j)D(\lambda_j) 
  = A(\lambda_j)D(\lambda_j) - B(\lambda_j)C(\lambda_j) 
  = Q(\lambda_j).  
  \label{eq:AD=Q-at-lj}
\eeq
Thus $A(\lambda_j) - D(\lambda_j)$ turns out to 
be expressed as 
\beq
  A(\lambda_j) - D(\lambda_j) 
  = \sqrt{P(\lambda_j)^2 - 4Q(\lambda_j)}. 
\eeq
This leads to the differential equation 
\beq
  \dot{\lambda}_j  
  = \frac{\sqrt{P(\lambda_j)^2 - 4Q(\lambda_j)}}{B'(\lambda_j)}  
  \label{eq:Dub}
\eeq
for the zeroes of $B(\lambda)$.  

The last equation resembles the so called 
``Dubrovin equation'' in the theory of 
finite-band integration 
\cite{bib:Novikov,bib:Lax,bib:Mc-vM}.  
In fact, if $\alpha = 0$, this is indeed a variant 
of Dubrovin equation that can be solved by the usual 
algebro-geometric method. If  $\alpha \not= 0$, 
the situation drastically changes.  Namely, 
the coefficients of $P(\lambda)$, as well as $b_0$, 
are no longer constant.  $\Gamma$ itself is 
accordingly a dynamical object, for which 
(\ref{eq:dot-P}) plays the role of equations 
of motion.   

The roots $\lambda_1,\ldots,\lambda_g$ of 
$B(\lambda)$ comprise a half of the spectral 
Darboux coordinates.  The other half 
$z_1,\ldots,z_g$ are defined by 
\beq
  z_j = A(\lambda_j). 
  \label{eq:def-zj}
\eeq
The $g$-tuple of points $(\lambda_j,z_j)$ of 
the $(\lambda,z)$ plane sit on the spectral curve 
$\Gamma$, namely, satisfy the algebraic relations 
\beq
  z_j^2 - P(\lambda_j)z_j + Q(\lambda_j) = 0, 
  \label{eq:lj-zj-on} 
\eeq
as one can readily see from the triangularity 
\beqnn
  T(\lambda_j) 
  = \left(\begin{array}{cc}
    A(\lambda_j) & 0 \\
    C(\lambda_j) & D(\lambda_j) 
    \end{array}\right) 
\eeqnn
of $T(\lambda_j)$.  We can prove that these 
moving points on $\Gamma$ satisfy the following 
differential equations.

\begin{prop}
$\lambda_j$ and $z_j$ satisfy the differential equations 
\beq
  \dot{\lambda}_j = \frac{z_j - Q(\lambda_j)z_j^{-1}}{B'(\lambda_j)}, 
  \label{eq:dot-lj} \\
  \dot{z}_j = \frac{P'(\lambda_j)z_j - Q'(\lambda_j)}{B'(\lambda_j)}. 
  \label{eq:dot-zj}
\eeq
\end{prop}

\proof
(\ref{eq:dot-lj}) is nothing but (\ref{eq:pre-Dub}) 
in disguise; the two terms in the denominator 
of (\ref{eq:pre-Dub}) can be rewritten as 
\beqnn
  A(\lambda_j) = z_j, \; 
  D(\lambda_j) = \frac{Q(\lambda_j)}{A(\lambda_j)} 
               = Q(\lambda_j)z_j^{-1}. 
\eeqnn
To derive (\ref{eq:dot-zj}), we differentiate 
(\ref{eq:lj-zj-on}) by $x$.  This yields 
\beqnn
  (2z_j - P'(\lambda_j))\dot{z}_j 
  + (- P'(\lambda_j)z_j + Q'(\lambda_j))\dot{\lambda}_j 
  - \dot{P}(\lambda_j)z_j = 0. 
\eeqnn
Solving this equation for $\dot{z}_j$ and noting 
the algebraic relation $\dot{P}(\lambda_j) = 0$ 
that holds by (\ref{eq:dot-P}), we find the equation 
\beqnn
  \dot{z}_j 
  = \frac{P'(\lambda_j)z_j - Q'(\lambda_j)}{2z_j - P(\lambda_j)} 
    \dot{\lambda}_j. 
\eeqnn
On the other hand, (\ref{eq:pre-Dub}) can also 
be written as 
\beqnn
  \dot{\lambda}_j 
  = \frac{2A(\lambda_j) - P(\lambda_j)}{B'(\lambda_j)} 
  = \frac{2z_j - P(\lambda_j)}{B'(\lambda_j)}. 
\eeqnn
Eliminating $\dot{\lambda}_j$ from these two equations, 
we obtain (\ref{eq:dot-zj}). 
\qed

\subsection{Deriving Hamiltonian system} 

We now seek to convert the differential equations 
of $\lambda_j$ and $z_j$ to a Hamiltonian system 
(which we call the ``spectral Hamiltonian system'').  
Since the situation is parallel to the case 
of the periodic Toda chains \cite{bib:Fl-Mc}, 
we first review that case, then turn to the case 
of periodic dressing chains.

\subsubsection{Periodic Toda chains}

The description of periodic Toda chains, too, starts 
from a $2 \times 2$ Lax pair and an associated 
transition matrix.  The spectral curve of an $N$-periodic 
chain is a hyperelliptic curve of genus $g = N-1$ defined 
by the equation 
\beq
  z^2 - P(\lambda)z +  Q = 0, 
\eeq
where $P(\lambda)$ and $Q$ are the trace and the 
determinant of the transition matrix.  Whereas 
$P(\lambda)$ is a polynomial of the form 
\beq
  P(\lambda) 
  = \lambda^N + \sum_{\ell=1}^N I_\ell\lambda^{N-\ell},  
\eeq
$Q$ is now a constant.  The next-to-leading term $I_1$ 
vanishes in the center-of-mass frame, and the other 
coefficients $I_2,\cdots,I_N$ are constants of motion; 
$I_2$ can be identified with the Toda Hamiltonian $H$.  
The first two matrix elements $A(\lambda)$ and $B(\lambda)$ 
of the transition matrix determine a set of moving points 
$(\lambda_1,z_1), \cdots, (\lambda_{N-1},z_{N-1})$ on 
the spectral curve as $B(\lambda_j) = 0$ and 
$z_j = A(\lambda_j)$.  Consequently, the coordinates 
of these moving points satisfy the algebraic equations 
\beq
  z_j^2 - P(\lambda_j)z_j + Q = 0. 
  \label{eq:lj-zj-on-Toda}
\eeq
Using the inversion formula (\ref{eq:Lag-v2}) 
in Appendix A, one can solve these equations for 
$I_\ell$ as 
\beq
  I_\ell 
  = - \sum_{j=1}^{N-1} 
      \frac{z_j + Qz_j^{-1} - \lambda_j^N}{B'(\lambda_j)}
      \frac{\rd b_\ell}{\rd \lambda_j}, 
\eeq
where $b_\ell$ are the coefficients of $B(\lambda)$, 
\beqnn
  B(\lambda) 
  = \lambda^{N-1} + b_1\lambda^{N-2} + \cdots + b_{N-1}, 
\eeqnn
the leading coefficient being equal to 
$1$ in the usual setup.  Since 
$b_1 = - \lambda_1 - \cdots - \lambda_{N-1}$, 
we have $\rd b_1/\rd \lambda_j = - 1$, 
so that 
\beq
  H = I_2 
    = \sum_{j=1}^{N-1} 
      \frac{z_j + Qz_j^{-1} - \lambda_j^N}{B'(\lambda_j)}. 
\eeq
The coordinates $(\lambda_j,z_j)$ of the moving 
points turn out to satisfy the Hamiltonian system 
\beq
  \dot{\lambda}_j = z_j \frac{\rd H}{\rd z_j}, \; 
  \dot{z}_j = - z_j \frac{\rd H}{\rd \lambda_j}.  
\eeq
Strictly speaking, it is $\lambda_j$ and $\log z_j$ 
that play the role of Darboux coordinates.  

The algebraic relations (\ref{eq:lj-zj-on-Toda}) 
imply that this Hamiltonian system is {\it separated} 
to a collection of independent systems on 
a two-dimensional phase space with coordinates 
$(\lambda_j,z_j)$; this is the central concept 
in ``separation of variables'' 
\cite{bib:Moser1,bib:AHH,bib:Sklyanin}. 
An integrable Hamiltonian system with a maximal 
number of constants of motion $I_1,\cdots,I_n$ 
in involution is {\it separable} if it has 
a suitable set of Darboux coordinates 
$\lambda_j,\mu_j$ that satisfy a set of 
equations of the form 
\beqnn
  f_j(\lambda_j,\mu_j,I_1,\cdots,I_n) = 0 
  \quad (j = 1,\cdots,n). 
\eeqnn
Geometrically, these equations define 
a family of Liouville tori, which are 
thus ``separated'' to a direct product of 
one-dimensional level curves.  For most 
integrable Hamiltonian systems with 
a Lax pair, these equations take 
an identical form 
\beqnn
  f(\lambda_j,\mu_j,I_1,\cdots,I_n) = 0 
  \quad (j=1,\cdots,n), 
\eeqnn
which stems from the the equation 
\beqnn
  f(\lambda,\mu,I_1,\cdots,I_n) = 0 
\eeqnn
of an associated spectral curve, 
as we have just illustrated for 
periodic Toda chains.

\subsubsection{Periodic dressing chains}

Bearing in mind this description of periodic Toda chains, 
we now turn to periodic dressing chains.  The algebraic 
equations (\ref{eq:lj-zj-on}) can be solved for $I_\ell$'s, 
$\ell = 1,\cdots,g$, as 
\beq
  I_\ell 
  = - \sum_{j=1}^g 
      \frac{z_j + Q(\lambda_j)z_j^{-1} - I_0\lambda_j^g}
           {B'(\lambda_j)}
      \frac{\rd b_\ell}{\rd \lambda_j} 
  \label{eq:Iell-Nodd}
\eeq
for $N = 2g + 1$, and 
\beq
  I_\ell
  = - \sum_{j=1}^g 
      \frac{z_j + Q(\lambda_j)z_j^{-1} - 2\lambda_j^{g+1} 
            - I_0\lambda_j^g}{B'(\lambda_j)} 
      \frac{\rd b_\ell}{\rd \lambda_j} 
  \label{eq:Iell-Neven}
\eeq
for $N = 2g + 2$.  Each $I_\ell$ thus becomes a function 
of the $2g$ variables $\lambda_1,\cdots,\lambda_g$ 
and $z_1,\cdots,z_g$ (as well as of $x$, through 
the $x$-dependence of $b_0$ and $I_0$, provided that 
$\alpha \not= 0$).  In view of the case of periodic 
Toda chains, it seems likely that the first nontrivial 
coefficient $I_1$ will play the role of the Hamiltonian.  
We have indeed the following result, which says that 
the analogy is essentially valid up to a simple factor.  
Note, however, that the Hamiltonian system thus obtained 
is {\it non-autonomous} (and presumably {\it not separable}) 
in the case where $\alpha \not= 0$.  

\begin{prop}\label{prop:Ham}
(\ref{eq:dot-lj}) and (\ref{eq:dot-zj}) can be cast 
into the Hamiltonian system 
\beq
  \dot{\lambda}_j = z_j \frac{\rd H}{\rd z_j}, \; 
  \dot{z}_j = - z_j \frac{\rd H}{\rd \lambda_j}, 
  \label{eq:Ham}
\eeq
where
\beq
  H = \frac{I_1}{b_0}. 
\eeq
More explicitly, the Hamiltonian takes the form 
\beq
  H = \sum_{j=1}^g 
      \frac{z_j + Q(\lambda_j)z_j^{-1} - I_0\lambda_j^g}
           {B'(\lambda_j)} 
\eeq
for $N = 2g + 1$ and 
\beq
  H = \sum_{j=1}^g 
      \frac{z_j + Q(\lambda_j)z_j^{-1} - 2\lambda_j^{g+1} 
            - I_0\lambda_j^g}{B'(\lambda_j)} 
\eeq
for $N = 2g + 2$.  
\end{prop}

\proof
It is obvious from the explicit form of $H$ that 
\beqnn
  z_j\frac{\rd H}{\rd z_j} 
  = \frac{z_j - Q(\lambda_j)z_j^{-1}}{B'(\lambda_j)}. 
\eeqnn
The first equation of (\ref{eq:Ham}) thus turns out 
to be the same thing as (\ref{eq:dot-lj}).  
To derive the second equation of (\ref{eq:Ham}), 
we notice that the equations 
\beqnn
  z_k^2 - P(\lambda_k)z_k + Q(\lambda_k) = 0 
\eeqnn
are identically satisfied if $I_\ell$ are understood 
to be a function of $\lambda_j$'s and $z_j$'s defined 
by (\ref{eq:Iell-Nodd}) or (\ref{eq:Iell-Neven}).  
Differentiating these equations by $\lambda_j$ 
yields the equations 
\beqnn
  (- P'(\lambda_k)z_k + Q'(\lambda_k))\delta_{jk} 
  - \sum_{\ell=1}^g 
    \frac{\rd I_\ell}{\rd \lambda_j}
    \lambda_k^{g-\ell}z_k = 0. 
\eeqnn
We can solve these equations for 
$\rd I_j/\rd \lambda_j$ by the inversion formula 
(\ref{eq:Lag-v2}) as 
\beqnn
  \frac{\rd I_\ell}{\rd \lambda_j} 
  &=& - \sum_{k=1}^g 
        \frac{(- P'(\lambda_j)z_k + Q'(\lambda_j))\delta_{jk}}
             {B'(\lambda_k)z_k}
        \frac{\rd b_\ell}{\rd \lambda_k} 
  \nonumber \\
  &=& \frac{P'(\lambda_j)z_j - Q'(\lambda_j)}{B'(\lambda_j)z_j}
      \frac{\rd b_\ell}{\rd \lambda_j}. 
\eeqnn
Since $b_1 = - b_0(\lambda_1 + \cdots + \lambda_g)$, 
we have 
\beqnn
  \frac{\rd b_1}{\rd \lambda_j} = - b_0, 
\eeqnn
so that 
\beqnn
    - z_j \frac{\rd H}{\rd \lambda_j} 
  = - \frac{z_j}{b_0} \frac{\rd I_1}{\rd \lambda_j} 
  = \frac{P'(\lambda_j)z_j - Q'(\lambda_j)}{B'(\lambda_j)}.  
\eeqnn
The second equation of (\ref{eq:Ham}) thus 
turns out to take the form 
\beqnn
  \dot{z}_j 
  = \frac{P'(\lambda_j)z_j - Q'(\lambda_j)}{B'(\lambda_j)}, 
\eeqnn
which is exactly (\ref{eq:dot-zj}).  
\qed

\subsection{Reconstructing Lax equation}

A significant feature of the spectral Hamiltonian system 
(\ref{eq:Ham}) is that it contains no information 
on $P(\lambda)$  (except for $I_0$).  In other words, 
this Hamiltonian system has apparently ``forgotten'' 
what the spectral curve is.  The only external input 
of the Hamiltonian are the polynomial $Q(\lambda)$ 
and the two quantities $b_0$ and $I_0$ that are 
a constant or a simple combination of $v$.  

Actually, alongside the spectral curve, one can 
reconstruct the transition matrix $T(\lambda)$ itself 
from the Hamiltonian system.  Thus the spectral 
Hamiltonian system is actually equivalent to 
the Lax equation (\ref{eq:dot-T}).

\subsubsection{If $N$ is odd} 

We first consider the case where $N = 2g + 1$.  
Since $b_0 = 1$, the external input of 
the Hamiltonian are $Q(\lambda)$ and $I_0 = 2v$; 
$v$ is assumed to be a function that satisfies 
(\ref{eq:dot-v}). 

The reconstruction proceeds as follows. 

i) Since $b_0 = 1$, we set 
\beq
  B(\lambda) 
  = \lambda^g + b_1\lambda^{g-1} + \cdots + b_g 
  = \prod_{j=1}^g (\lambda - \lambda_j), 
  \label{eq:re-B-Nodd}
\eeq
and define the polynomial 
\beq
  A(\lambda) = a_0 \lambda^g + a_1\lambda^{g-1} + \cdots + a_g, \; 
  a_0 = \frac{I_0}{2}, 
\eeq
by the equations 
\beq
  A(\lambda_j) = z_j  \quad (j = 1,\ldots,g).  
  \label{eq:re-A-Nodd}
\eeq

ii) We now define
\beq
  D(\lambda) &=& P(\lambda) - A(\lambda), 
  \label{eq:re-D-Nodd} \\
  C(\lambda) &=& \frac{A(\lambda)D(\lambda) - Q(\lambda)}{B(\lambda)}. 
  \label{eq:re-C-Nodd}
\eeq
$D(\lambda)$ is a polynomial of the form 
$D(\lambda) = d_0\lambda^g + d_1\lambda^{g-1} + \cdots + d_g$ 
with $d_0 = I_0 - a_0 = I_0/2$. $C(\lambda)$, too, 
becomes a polynomial, because the numerator vanishes 
at the zeroes $\lambda = \lambda_j$ of the denominator, 
see (\ref{eq:AD=Q-at-lj}).  
Since $A(\lambda)D(\lambda) - Q(\lambda)$ is 
a polynomial of degree $2g+1$ with the leading 
coefficients being equal to $1$, $C(\lambda)$ 
is a polynomial of the form 
$C(\lambda) = \lambda^{g+1} + c_1\lambda^g + \cdots + c_{g+1}$. 
The four matrix elements of $T(\lambda)$ have been 
thus determined.  

iii) Let us examine the first equation of (\ref{eq:Ham}).  
Calculating the right hand side explicitly, 
we can rewrite this equation as 
\beqnn
  B'(\lambda)\dot{\lambda}_j = A(\lambda_j) - D(\lambda_j). 
\eeqnn
By the identity 
$\dot{B}(\lambda_j) + B'(\lambda_j)\dot{\lambda}_j = 0$, 
the last equation can be further rewritten as 
\beqnn
  \dot{B}(\lambda_j) + A(\lambda_j) - D(\lambda_j) = 0. 
\eeqnn
On the other hand, since $b_0 = 1$ and $a_0 = d_0$, 
$\dot{B}(\lambda) + A(\lambda) - D(\lambda)$ is 
a polynomial of degree less than $g$.  Such a polynomial 
cannot vanish at $g$ distinct points unless it is 
identically zero.  We thus arrive at the equation 
\beqnn
  \dot{B}(\lambda) + A(\lambda) - D(\lambda) = 0, 
\eeqnn
which is exactly (\ref{eq:dot-B}).  

iv) Let us examine the second equation of (\ref{eq:Ham}).  
Viewing the algebraic relations 
$z_k^2 - P(\lambda_k)z_k + Q(\lambda_k) = 0$ 
as identities satisfied by $I_\ell$'s, we repeat 
the calculations in the proof of Proposition \ref{prop:Ham} 
and obtain the expression 
\beqnn
  - z_j\frac{\rd H}{\rd \lambda_j} 
  = \frac{P'(\lambda_j)z_j - Q'(\lambda_j)}{B'(\lambda_j)} 
\eeqnn
for the right hand side of the second equation 
of (\ref{eq:Ham}).  On the other hand, 
differentiating the same identities by $x$ 
along the trajectories of (\ref{eq:Ham}) gives 
\beqnn
  (2z_j - P'(\lambda_j))\dot{z}_j 
  + (- P'(\lambda_j)z_j + Q'(\lambda_j))\dot{\lambda}_j 
  - \dot{P}(\lambda_j)z_j = 0, 
\eeqnn
which implies 
\beqnn
  \dot{z}_j 
  &=& \frac{P'(\lambda_j)z_j - Q'(\lambda_j)}{2z_j - P(\lambda_j)} 
      \dot{\lambda_j} 
    + \frac{\dot{P}(\lambda_j)}{2z_j - P(\lambda_j)}z_j 
  \\
  &=& \frac{P'(\lambda_j)z_j - Q'(\lambda_j)}{B'(\lambda_j)} 
    + \frac{\dot{P}(\lambda_j)}{2z_j - P(\lambda_j)}z_j. 
\eeqnn
For the second equation of (\ref{eq:Ham}) to hold, 
the algebraic relations $\dot{P}(\lambda_j) = 0$ 
($j = 1,\cdots,g$) have to be satisfied 
This implies that $\dot{P}(\lambda)$ is divisible 
by $B(\lambda)$.  Moreover, since $P(\lambda)$ is 
a polynomial of degree $g$ with the leading coefficient 
being equal to $\alpha$, we eventually obtain 
the equation 
\beqnn
  \frac{\dot{P}(\lambda)}{B(\lambda)} = \alpha, 
\eeqnn
which is nothing but (\ref{eq:dot-P}).  

v) In order to derive (\ref{eq:dot-A}), 
we start from the identity 
$\dot{A}(\lambda_j) + A'(\lambda_j)\dot{\lambda}_j = \dot{z}_j$ 
that can be obtained from (\ref{eq:re-A-Nodd}) .  
Since  $\dot{\lambda}_j$ and $\dot{z}_j$ can be 
calculated as shown in (\ref{eq:pre-Dub}) and 
(\ref{eq:dot-zj}), we have 
\beqnn
  \dot{A}(\lambda_j) 
  = \frac{D'(\lambda_j)A(\lambda_j) + A'(\lambda_j)D(\lambda_j) 
            - Q'(\lambda_j)}{B'(\lambda_j)}. 
\eeqnn
On the other hand, the value of $C(\lambda)$ 
at $\lambda = \lambda_j$ can be calculated 
by de L'Hopital's formula as 
\beqnn
  C(\lambda_j) 
  &=& \lim_{\lambda\to\lambda_j} 
      \frac{A(\lambda)D(\lambda) - Q'(\lambda)}{B(\lambda)} 
  \\
  &=& \frac{A'(\lambda_j)D(\lambda_j) + A(\lambda_j)D'(\lambda_j) 
            - Q'(\lambda_j)}{B'(\lambda_j)}. 
\eeqnn
Since the right hand side of these two equalities 
are identical, we find that 
\beqnn
  \dot{A}(\lambda_j) = C(\lambda_j) 
\eeqnn
for $j = 1,\cdots,g$, so that 
$\dot{A}(\lambda) - C(\lambda)$ is divisible by 
$B(\lambda)$.  On account of the Laurent expansion 
of the quotient at $\lambda = \infty$, we can deduce that 
\beqnn
  \frac{\dot{A}(\lambda) - C(\lambda)}{B(\lambda)} 
  = - \lambda + \dot{a}_0 - c_1 + b_1. 
\eeqnn
This equation coincides with 
(\ref{eq:dot-A}) upon identifying 
\beq
  u_1 = - \dot{a}_0 + c_1 - b_1. 
  \label{eq:re-u1-Nodd}
\eeq

vi) The final step is to verify (\ref{eq:dot-C}) and 
(\ref{eq:dot-D}).  Firstly, we now have (\ref{eq:dot-A}) 
and (\ref{eq:dot-P}) at hand, and using these equations 
we can derive (\ref{eq:dot-D}): 
\beqnn
  \dot{D}(\lambda) 
  = \dot{P}(\lambda) - \dot{A}(\lambda) 
  = (\lambda + u_1 + \alpha)B(\lambda) - C(\lambda). 
\eeqnn
To derive (\ref{eq:dot-C}), we differentiate the identity 
$Q(\lambda) = A(\lambda)D(\lambda) - B(\lambda)C(\lambda)$ 
by $x$ and eliminate the derivatives 
$\dot{A}(\lambda),\dot{B}(\lambda)$ and 
$\dot{D}(\lambda)$ using the differential 
equations that have been proven: 
\beqnn
  0 &=& \dot{A}(\lambda)D(\lambda) + A(\lambda)\dot{D}(\lambda) 
      - \dot{B}(\lambda)C(\lambda) - B(\lambda)\dot{C}(\lambda) \\ 
    &=& B(\lambda)\Bigl((\lambda + \alpha + u_1)A(\lambda) 
          - (\lambda + u_1)D(\lambda) - \dot{C}(\lambda)\Bigr).  
\eeqnn
Removing the common factor $B(\lambda)$, 
we obtain (\ref{eq:dot-C}).

\subsubsection{If $N$ is even} 

Let us now consider the case where $N = 2g + 2$.  
The external input of the Hamiltonian are 
$Q(\lambda)$, $b_0 = v$ and 
$I_0 = v^2 + \beta_1 + \cdots + \beta_{2g+1}$; 
$v$ is assumed to be a function that satisfies 
(\ref{eq:dot-v}). 

The reconstruction of $T(\lambda)$ is 
mostly parallel to the previous case. 
$B(\lambda)$ is the polynomial 
\beq
  B(\lambda) 
  = b_0\lambda^g + b_1\lambda^{g-1} + \cdots + b_g 
  = b_0 \prod_{j=1}^g (\lambda - \lambda_j). 
  \label{eq:re-B-Neven}
\eeq
$A(\lambda)$ is a polynomial of the form 
\beq
  A(\lambda) = \lambda^{g+1} + a_0\lambda^g + \cdots + a_g, \; 
  a_0 = \frac{1}{2}(I_0 - \dot{b}_0), 
\eeq
and defined by the equations 
\beq
  A(\lambda_j) = z_j  \quad (j=1,\cdots,g). 
\eeq
$C(\lambda)$ and $D(\lambda)$ are given by 
\beq
  D(\lambda) &=& P(\lambda) - A(\lambda), 
  \label{eq:re-D-Neven} \\
  C(\lambda) &=& \frac{A(\lambda)D(\lambda) - Q(\lambda)}{B(\lambda)}. 
  \label{eq:re-C-Neven} 
\eeq
They become polynomials of the form 
$D(\lambda) = \lambda^{g+1} + d_0\lambda^g + d_1\lambda^{g-1} 
+ \cdots + d_g$ and 
$C(\lambda) = c_0\lambda^{g+1} + c_1\lambda^g + \cdots + c_{g+1}$ 
with $d_0 = (I_0 + \dot{b}_0)/2$ and $c_0 = b_0$.  
As we have done for odd $N$'s, these four polynomials 
turn out to satisfy (\ref{eq:dot-A}), (\ref{eq:dot-B}), 
(\ref{eq:dot-C}) and (\ref{eq:dot-D}) with $u_1$ of 
the form 
\beq
  u_1 = \frac{- \dot{a}_0 + c_1 + b_1}{b_0}. 
  \label{eq:re-u1-Neven} 
\eeq
\bigskip

We have thus proven the following: 

\begin{prop}\label{prop:repro} 
Given a solution of the spectral Hamiltonian system 
(\ref{eq:Ham}), one can reconstruct the transition matrix 
$T(\lambda)$ and the potential $u_1$ by 
(\ref{eq:re-B-Nodd}) -- (\ref{eq:re-u1-Nodd}) 
(for $N = 2g + 1$) and 
(\ref{eq:re-B-Neven}) -- (\ref{eq:re-u1-Neven}) 
(for $N = 2g + 2$).  
\end{prop}

The contents of this section can be summarized as follows. 

\begin{theorem}
The Lax equation (\ref{eq:dot-T}) of 
the transition matrix $T(\lambda)$ is equivalent 
to the spectral Hamiltonian system (\ref{eq:Ham}) 
in the spectral Darboux coordinates. 
\end{theorem}

\newsection{Inverse problem and Poisson structures}

We now turn to the ``inverse problem'', namely, 
the problem of reconstructing the phase space 
coordinates of the periodic dressing chain, or 
of the Noumi-Yamada system, from the spectral 
Darboux coordinates.  We can solve this problem 
and find that the spectral Hamiltonian system 
and the Noumi-Yamada system are connected 
by a locally invertible rational map 
$(\lambda_1,\ldots,\lambda_g,z_1,\ldots,z_g,v) 
\mapsto (f_1,\ldots,f_N)$.  Moreover, this map 
turns out to be a Poisson map connecting 
two odd-dimensional Poisson structures.

\subsection{General setup and case study}

The phase space of the spectral Hamiltonian system 
has a natural symplectic (hence, even-dimensional 
Poisson) structure.  Actually, one can extend 
this phase space to an odd-dimensional 
Poisson manifold with an extra dimension. 
On the other hand, the phase space of 
the Noumi-Yamada system is known to have 
an odd-dimensional Poisson structure 
\cite{bib:Ve-Sh,bib:No-Ya1}.  

The two Poisson structures are formulated as follows. 
\begin{enumerate}
\item 
The first Poisson structure is defined on 
the {\it extended} phase space of the spectral 
Hamiltonian system (\ref{eq:Ham}) with 
the auxiliary variable $v$ added as 
an extra dimension.   This is a $(2g +1)$-dimensional 
Poisson manifold with the Poisson brackets 
\beq
  \{\lambda_j,z_k\} = \delta_{jk}, \; 
  \{\lambda_j,\lambda_k\} = \{z_j,z_k\} 
  = \{\lambda_j,v\} = \{z_j,v\} = 0. 
  \label{eq:PB-lj-zj-v}
\eeq
In particular, $v$ is a Casimir function, i.e., 
a central element of the Poisson algebra of 
coordinates.  
\item 
The second Poisson structure is defined on the phase space 
of the Noumi-Yamada system endowed with the Poisson brackets 
\beq
  \{f_m, f_n\} = - \delta_{m+1,n} + \delta_{m-1,n}. 
  \label{eq:PB-fn}
\eeq
Since the variables $f_n$ are required to satisfy 
the constraint (\ref{eq:fn-constraint}) ,  
the phase space in this case, too, is 
$(2g + 1)$-dimensional.  The auxiliary function 
$v = (f_1 + \cdots + f_N)/2$ is a Casimir function 
of this Poisson structure as well.  
\end{enumerate}

Let us now examine the simplest two cases, 
i.e, $N = 3$ and $N = 4$.

\subsubsection{Case study: $N = 3$} 

The three-periodic dressing chain has the three fundamental 
variables $v_1,v_2,v_3$, which are connected with the 
variables of the Noumi-Yamada system of $A^{(1)}_2$ type: 
\beq
  f_1 = v_1 + v_2, \; f_2 = v_2 + v_3, \; f_3 = v_3 + v_1. 
\eeq
The phase space of the Noumi-Yamada system is now a 
three-dimensional Poisson manifold with the Poisson brackets 
\beq
  \{f_1,f_2\} = \{f_2,f_3\} = \{f_3,f_1\} = - 1 
\eeq
or, equivalently, 
\beq
  \{v_1,v_2\} = -1,\; \{v_1,v_3\} = 1, \; \{v_2,v_3\} = -1. 
\eeq
It is easy to verify that the auxiliary function 
$v = v_1 + v_2 + v_3 = (f_1 + f_2 + f_3)/$ is indeed 
a Casimir function of this Poisson structure.  
This is the Poisson structure that appears, 
along with a generalization to odd $N$'s, 
in the work of Veselov and Shabat \cite{bib:Ve-Sh}.  

The equation of motion (\ref{eq:NY-Nodd}) can be 
rewritten as 
\beq
  \dot{f}_1 = \{f_1, H\}, \; \dot{f}_2 = \{f_2, H\}, \; 
  \dot{f}_3 = \{f_3, H\} + \alpha, 
\eeq
where $H$ is the function that can be calculated from 
the transition matrix as 
\beq
  H = \Tr T(0) = f_1f_2f_3 + \beta_1f_3 + \beta_2f_1.  
\eeq
Actually, this is the same Hamiltonian as that of 
(\ref{eq:Ham}), i.e., $H = I_1$.  This ``Liouville form'' 
of the equation of motion is derived from a quite different 
point of view in the work of Noumi and Yamada \cite{bib:No-Ya1} 

The construction of the spectral Darboux coordinates 
now takes a particularly simple form, because $B(\lambda)$ 
and $A(\lambda)$ are a linear function of $\lambda$: 
\beq
  B(\lambda) = \lambda + f_1 f_2 + \beta_1, \; 
  A(\lambda) = v \lambda + (f_1 f_2 + \beta_1)v_1. 
\eeq
The spectral Darboux coordinates $\lambda_1,z_1$ are 
thus determined explicitly as 
\beq
  \lambda_1 = - f_1 f_2 - \beta_1, \quad 
  z_1 = f_2 \lambda_1. 
\eeq
Note that this expression of $\lambda_1$ and $z_1$ 
can be solved for $f_1$ and $f_2$ as 
\beq
  f_2 = \frac{z_1}{\lambda_1}, \; 
  f_1 = - \frac{\lambda_1(\lambda_1 + \beta_1)}{z_1}, 
\eeq
so that, along with $f_3 = v - f_1 -f_2$, all $f_n$'s 
are a rational function of $\lambda_1,z_1$ and $v$.  
Moreover, one can easily verify that the Poisson brackets 
of $f_n$ are consistent with the Poisson brackets 
\beq
  \{\lambda_1,z_1\} = z_1, \; 
  \{\lambda_1,v\} = \{z_1,v\} = 0 
\eeq
of $\lambda_1,z_1$ and $v$.  In other words, 
the rational map $(f_1,f_2,f_3) \mapsto (\lambda_1,z_1,v)$ 
is Poisson.  

As a final remark, let us point out that $A(\lambda)$ 
can be also expressed as 
\beq
  A(\lambda) = B(\lambda)v_1 + f_2\lambda. 
\eeq
We shall encounter a similar expression in the case study 
for $N = 4$ below.

\subsubsection{Case study: $N = 4$} 

The four-periodic dressing chain has the four fundamental 
variables $v_1,\ldots,v_4$ under the quadratic constraint 
(\ref{eq:vn-quad-constr}).  The four variables $f_1,\ldots,f_4$ 
of the Noumi-Yamada system of $A^{(1)}_3$ type are defined 
by the linear relations 
\beq
  f_1 = v_1 + v_2, \; f_2 = v_2 + v_3, \; 
  f_3 = v_3 + v_4, \; f_4 = v_4 + v_1, 
\eeq
and obey the constraint $f_1 + f_3 = f_2 + f_4$.  
They are thus a set of (redundant) coordinates of 
a three-dimensional Poisson manifold with the Poisson brackets 
\beq
  \{f_1,f_2\} = \{f_2,f_3\} = \{f_3,f_4\} = \{f_4,f_1\} = -1, 
  \nonumber \\
  \{f_1,f_3\} = \{f_1,f_4\} = \{f_2,f_4\} = 0. 
\eeq
The auxiliary variable $v = f_1 + f_3 = f_2 + f_4$ is 
a Casimir function.  

As we have seen in Section 2, the linear relations 
connecting $f_n$'s with $v_n$'s can be solved for $v_n$'s 
under the constraint (\ref{eq:vn-quad-constr}).  
The outcome is a rational expression of $v_n$'s 
in terms of $f_n$'s, such as 
\beq
  v_1 
  = \frac{1}{4v}(2v^2 - 4f_2f_3 + \alpha_1 - \alpha_2 
      + \alpha_3 - \alpha_4), 
  \label{eq:v1-fn-N=4}
\eeq
etc.  The Poisson brackets of $v_n$'s are thereby 
determined by the Poisson brackets of $f_n$'s as 
\beq
  \{v_1,v_2\} = - \frac{f_3}{v}, \; 
  \{v_1,v_3\} = - \frac{f_2}{v} + \frac{f_3}{v}, \; 
  \{v_1,v_4\} = \frac{f_2}{v}, \nonumber \\
  \{v_2,v_3\} = - \frac{f_4}{v}, \; 
  \{v_2,v_4\} = - \frac{f_3}{v} + \frac{f_4}{v}, \; 
  \{v_3,v_4\} = - \frac{f_1}{v}. 
\eeq
For example, the Poisson bracket of $v_1$ and $v_2$ 
can be correctly derived from the identity $\{v_1,v_2\} = 
\{v_1, f_1 - v_1\} = \{v_1,f_1\}$ and (\ref{eq:v1-fn-N=4}). 

The equation of motion (\ref{eq:NY-Neven}) again turns out 
to take Liouville form 
\beq
  \dot{f}_1 = \{f_1, H\}, \; 
  \dot{f}_2 = \{f_2, H\}, \; 
  \dot{f}_3 = \{f_3, H\}, \; 
  \dot{f}_4 = \{f_4, H\} + \alpha 
\eeq
with the Hamiltonian 
\beq
  H = v^{-1} \Tr T(0)
    = v^{-1}(f_1 f_2 f_3 f_4 + \beta_1 f_3 f_4 
         + \beta_2 f_1 f_4 + \beta_3 f_1 f_2 
         + \beta_1 \beta_3). 
\eeq
This Hamiltonian $H$ is the same thing that appears 
in (\ref{eq:Ham}), i.e., $H = I_1/v$.  

The construction of the spectral Darboux coordinates 
is mostly parallel to the case of $N = 3$, because 
$B(\lambda)$ is a linear function of $\lambda$.  
More precisely, we have 
\beq
  B(\lambda) 
  = v \lambda + f_1 f_2 f_3 + \beta_1 f_3 + \beta_2 f_1, 
  \nonumber \\
  A(\lambda) 
  = B(\lambda)v_1 + (f_2 f_3 + \lambda + \beta_2)\lambda. 
\eeq
Note that we have expressed $A(\lambda)$ in a particular form, 
which we shall generalize to all $N$'s.  The spectral 
Darboux coordinates $\lambda_1,z_1$ are thus determined as 
\beq
  \lambda_1 = - v^{-1}(f_1 f_2 f_3 + \beta_1 f_3 + \beta_2 f_1), \; 
  z_1 = (f_2 f_3 + \lambda_1 + \beta_2)\lambda_1. 
\eeq
Combined with the linear relations $f_1 + f_3 = f_2 + f_4 = v$, 
these formulae of $\lambda_1,z_1$ can be solved for $f_n$'s.  
This yields an expression of $f_n$'s as a rational function 
of $\lambda_1,z_1$ and $v$, which can be assembled to 
a rational map $(\lambda_1,z_1,v) \mapsto (f_1,f_2,f_3,f_4)$.  
This map, too, turns out to be a Poisson map with regard to 
the aforementioned Poisson structures.

\subsection{Solving inverse problem}

Let us now consider, in general, how to reconstruct $v_n$'s 
or $f_n$'s from the spectral Darboux coordinates $\lambda_j,z_j$ 
and the auxiliary variable $v$.  Since the latter carry 
the same information as the transfer matrix $T(\lambda)$, 
the problem is to reconstruct $v_n$'s or $f_n$'s from 
the transfer matrix.  

To solve this problem, we make use of a new auxiliary 
polynomial $\Atilde(\lambda)$ such that 
\beq
  A(\lambda) = B(\lambda)v_1 + \Atilde(\lambda)\lambda. 
  \label{eq:A-Atilde}
\eeq
This polynomial and $B(\lambda)$ have several interesting 
properties, which eventually lead to a solution of 
the inverse problem.

\subsubsection{Structure of $\Atilde(\lambda)$ and $B(\lambda)$}

The existence of the polynomial $\Atilde(\lambda)$ 
has been hinted in the case study for $N = 3$ and $N = 4$.  
Not only the existence for all $N \ge 3$, we can also 
deduce the following result on its precise structure.  
Note, in particular, that $\Atilde(\lambda)$ and $B(\lambda)$ 
turn out to be a function of $f_n$'s rather than $v_n$'s.  

\begin{prop}
For all $N \ge 3$, there is a polynomial $\Atilde(\lambda)$ 
that satisfies (\ref{eq:A-Atilde}).  Moreover, 
this polynomial can be written as 
\beq
  \Atilde(\lambda) 
  &=& \prod_{n=2}^{N-2} \left(
      1 + (\lambda + \beta_n)
          \frac{\rd^2}{\rd f_n \rd f_{n+1}}
      \right)(f_2 \cdots f_{N-1}) 
  \label{eq:Atilde-fn} 
\eeq
for $N \ge 4$ and $\Atilde(\lambda) = f_2$ 
for $N = 3$.  Moreover, $B(\lambda)$, too, can be 
similarly expressed as 
\beq
  B(\lambda) 
  &=& \prod_{n=1}^{N-2} \left(
      1 + (\lambda + \beta_n)
          \frac{\rd^2}{\rd f_n \rd f_{n+1}}
      \right)(f_1 \cdots f_{N-1}). 
  \label{eq:B-fn} 
\eeq
\end{prop}

\proof
Let us note that $A(\lambda)$ and $B(\lambda)$ can 
be expressed as 
\beqnn
  A(\lambda) 
  &=& \Tr 
    \left(\begin{array}{cc} 1 & 0 \end{array}\right) 
    V_N(\lambda + \beta_{N-1}) \cdots V_1(\lambda) 
    \left(\begin{array}{c} 
    1 \\
    0 
    \end{array}\right) 
  \\ 
  &=& \Tr 
    \left(\begin{array}{cc} 1 & 0 \end{array}\right) 
    V_N(\lambda + \beta_{N-1}) \cdots V_2(\lambda + \beta_1) 
    \left(\begin{array}{c}
    v_1 \\
    \lambda + v_1^2 
    \end{array}\right). 
\eeqnn
and 
\beqnn
  B(\lambda) 
  &=& \Tr 
    \left(\begin{array}{cc} 1 & 0 \end{array}\right) 
    V_N(\lambda + \beta_{N-1}) \cdots V_1(\lambda) 
    \left(\begin{array}{c} 
    0 \\
    1 
    \end{array}\right) 
  \nonumber \\ 
  &=& \Tr 
    \left(\begin{array}{cc} 1 & 0 \end{array}\right) 
    V_N(\lambda + \beta_{N-1}) \cdots V_2(\lambda + \beta_1) 
    \left(\begin{array}{c}
    1 \\
    v_1 
    \end{array}\right). 
\eeqnn
Substituting 
\beqnn
    \left(\begin{array}{c}
    v_1 \\
    \lambda + v_1^2 
    \end{array}\right) 
  = v_1 
    \left(\begin{array}{c}
    1 \\
    v_1 
    \end{array}\right) 
  + \lambda 
    \left(\begin{array}{c}
    0 \\
    1 
    \end{array}\right), 
\eeqnn
we can rewrite the last expression of $A(\lambda)$ as 
\beqnn
  A(\lambda) 
  &=& \Tr \left(\begin{array}{cc} 1 & 0 \end{array}\right) 
      V_N(\lambda + \beta_{N-1}) \cdots V_2(\lambda + \beta_1) 
      \left(\begin{array}{c} 
      1 \\ 
      v_1 
      \end{array}\right) v_1 
  \\
  && \mbox{} 
    + \Tr \left(\begin{array}{cc} 1 & 0 \end{array}\right) 
      V_N(\lambda + \beta_{N-1}) \cdots V_3(\lambda + \beta_2) 
      \left(\begin{array}{c} 
      1 \\ 
      v_2  
      \end{array}\right) \lambda. 
\eeqnn
Since the first term on the right hand side 
is equal to $B(\lambda) v_1$, one obtains 
(\ref{eq:A-Atilde}) upon defining 
$\Atilde(\lambda)$ as 
\beqnn
  \Atilde(\lambda) 
  = \Tr \left(\begin{array}{cc} 1 & 0 \end{array}\right) 
    V_N(\lambda + \beta_{N-1}) \cdots V_3(\lambda + \beta_2) 
    \left(\begin{array}{c} 
    1 \\ 
    v_2  
    \end{array}\right).   
\eeqnn
We can now apply the method used for evaluating 
the trace formula (\ref{eq:P-as-trace}) to 
the right hand side of these formulae.  
This yields (\ref{eq:Atilde-fn}) and 
(\ref{eq:B-fn}).   \qed

\subsubsection{Three-term recursion relations} 

The foregoing expression of $\Atilde(\lambda)$ 
and $B(\lambda)$ shows that they are the first two 
members of the finite sequence of polynomials 
\beq
  F_m(\lambda) 
  = \prod_{n=m}^{N-2} \left(
      1 + (\lambda + \beta_n)
          \frac{\rd^2}{\rd f_n \rd f_{n+1}}
    \right)(f_m \cdots f_{N-1}) 
  \quad (m = 1,\ldots,N-2)
\eeq
and $F_{N-1}(\lambda) = f_{N-1}$.  
An interesting property of these polynomials is 
that they satisfy a set of three-term recursion relations 
as we show below.   It deserves to be mentioned that 
a similar finite sequence of polynomials take place 
in Moser's method for solving finite non-periodic 
Toda chains \cite{bib:Moser2}.  

\begin{prop}
The polynomials $F_m(\lambda)$ satisfy the 
three-term linear recursion relations 
\beq
  F_m(\lambda) = f_m F_{m+1}(\lambda) 
    + (\lambda + \beta_m)F_{m+2}(\lambda) 
  \label{eq:3-term-relation}
\eeq
and the auxiliary relation 
\beq
  \frac{\rd F_m(\lambda)}{\rd f_m} = F_{m+1}(\lambda). 
  \label{eq:Fm-F(m+1)}
\eeq
\end{prop}

\proof 
The definition of $F_m(\lambda)$ can be rewritten as 
\beqnn
  F_m(\lambda) 
  &=& \left(1 + (\lambda + \beta_m)
                \frac{\rd^2}{\rd f_m \rd f_{m+1}}
      \right)(f_m F_{m+1}(\lambda)) \\
  &=& f_m F_{m+1}(\lambda) 
    + (\lambda + \beta_m) 
      \frac{\rd F_{m+1}(\lambda)}{\rd f_{m+1}}. 
\eeqnn
It is also obvious from the construction that 
$F_{m+1}(\lambda)$ does not contain $f_m$.  
Since $f_m$ appears only in the first term 
on the right hand side, one immediately finds 
that (\ref{eq:Fm-F(m+1)}) holds.  The last equality 
thereby turns into the three-term linear recursion relation.  
\qed 

An immediate consequence of the three-term recursion 
relations is that $F_m(\lambda)$  has the determinant 
formula 
\beq
  F_m(\lambda) 
  = \left|\begin{array}{cccc}
    f_m              &1       &      &  \\
    -\lambda-\beta_m &f_{m+1} &\ddots&  \\
                     &\ddots  &\ddots&1 \\
                     &        &-\lambda-\beta_{N-2}&f_{N-1}
    \end{array}\right|. 
\eeq
This formula shows that the polynomials $F_m(\lambda)$ 
are also closely related to the Lax pair of 
Noumi and Yamada.  It is easy to prove this formula:  
Firstly, one can easily verify that these determinants 
satisfy the same recursion relations. Secondly, 
the formula holds for $m = N-2$ and $m = N-1$.  
Consequently, the determinant has to coincide 
with $F_m(\lambda)$ for all $m$'s.

\subsubsection{Reconstructing $f_n$'s} 

We are now in a position to solve the problem 
of reconstructing $f_n$'s from the spectral 
Darboux coordinates $\lambda_j,z_j$ and 
the auxiliary variable $v$.  A clue is the relation 
\beq
  F_m(-\beta_m) = f_m F_{m+1}(-\beta_m) 
\eeq
that can be derived from the three-term recursion relations.  
This relation may be thought of as a formula that determines 
$f_m$ from $F_m(\lambda)$ and $F_{m+1}(\lambda)$.  Similarly, 
$F_{m+2}(\lambda)$ can be expressed as 
\beq
  F_{m+2}(\lambda) = 
    \frac{F_m(\lambda) - f_mF_{m+1}(\lambda)}{\lambda + \beta_m}. 
\eeq
Thus $f_m$ and $F_{m+2}(\lambda)$ can be recursively 
determined from $F_1(\lambda) = B(\lambda)$ and 
$F_2(\lambda) = \Atilde(\lambda)$.  
As regards $B(\lambda)$ and $\Atilde(\lambda)$, 
they are polynomials of the form 
\beqnn
  B(\lambda) = b_0\lambda^g + \cdots + b_g, \; 
  \Atilde(\lambda) = \atilde_0\lambda^g + \cdots + \atilde_g, 
\eeqnn
where $b_g = 1,\; \atilde_0 = 0$ for $N = 2g + 1$ 
and $b_0 = v,\; \atilde_0 = 1$ for $N = 2g + 2$, 
and satisfy the algebraic relations 
\beq
  B(\lambda_j) = 0, \; z_j = \Atilde(\lambda_j)\lambda_j 
  \label{eq:lj-zj-Atilde-B}
\eeq
as one can readily see from (\ref{eq:A-Atilde}).  
Consequently, by the interpolation formula (\ref{eq:Lag-mod}), 
\beq
  B(\lambda) = b_0 \prod_{j=1}^g(\lambda - \lambda_j), \quad 
  \Atilde(\lambda) 
  = \left(\frac{\atilde_0}{b_0}  
      + \sum_{j=1}^g \frac{z_j}
        {B'(\lambda_j)(\lambda - \lambda_j)\lambda_j}
    \right)B(\lambda). 
  \label{eq:Atilde-B-lj-zj}
\eeq
Starting from these polynomials, one can thus determine 
$f_n$'s successively as 
\beq
  f_1 = \frac{B(-\beta_1)}{\Atilde(-\beta_1)}, \quad 
  f_2 = \frac{\Atilde(-\beta_2)(-\beta_2 + \beta_1)}
        {B(-\beta_2) - f_1\Atilde(-\beta_2)}, \quad   
  \mbox{etc.}
  \label{eq:fn-lj-zj-v}
\eeq
In particular, each $f_n$ is a rational function of 
the spectral Darboux coordinates and $v$.  

This completes the reconstruction of $f_n$'s 
in the case where $N = 2g + 1$.  If $N = 2g + 2$, 
we have to confirm that $f_n$'s satisfy the constraint 
(\ref{eq:fn-constraint}) as well.  Actually, 
this is an immediate consequence of the recursive 
formulae of $F_m(\lambda)$'s:  Starting from 
the aforementioned form of $F_1(\lambda) = B(\lambda)$ 
and $F_2(\lambda) = \Atilde(\lambda)$, one can show 
by induction that 
\beq
  F_{2n}(\lambda) &=& \lambda^{g-n+1} 
    + \mbox{terms of lower degree}, 
  \nonumber \\
  F_{2n+1}(\lambda) &=& 
    (v - f_1 - f_3 - \cdots - f_{2n-1})\lambda^{g-n} 
    + \mbox{terms of lower degree} 
\eeq
for $n = 1,2,\ldots,g$, so that, eventually, 
\beqnn
  f_{2g+1} = F_{2g+1}(\lambda) 
           = v - f_1 - f_3 - \ldots - f_{2g-1}. 
\eeqnn
This implies (\ref{eq:fn-constraint}).  

We have thus proven the following.  

\begin{theorem}
The Noumi-Yamada system is connected with 
the spectral Hamiltonian system by 
a locally invertible rational map 
$(\lambda_1,\ldots,\lambda_g,z_1,\ldots,z_g,v) 
\mapsto (f_1,\ldots,f_N)$.  This map consists 
of rational functions recursively calculated 
as (\ref{eq:fn-lj-zj-v}).  
\end{theorem}

\subsection{Comparing two Poisson structure}

Now that the spectral Hamiltonian system and 
the Noumi-Yamada system are connected by 
a locally invertible map, one can compare 
the Poisson structures on the (extended) 
phase spaces of these systems.  Actually, 
it is more convenient to consider 
this issue at a place in between, namely, 
on the $(2g + 1)$-dimensional space with 
coordinates $(\atilde_1,\ldots,\atilde_g,
b_1,\ldots,b_g,v)$.  

If one reviews the foregoing construction carefully, 
one will soon notice that the map constructed above 
can be factorized to the composition of two maps as 
\beq
  (\lambda_1,\ldots,\lambda_g,z_1,\ldots,z_g,v) 
  \stackrel{\gamma_1}{\longmapsto} 
  (\atilde_1,\ldots,\atilde_g,b_1,\ldots,b_g,v) 
  \stackrel{\gamma_2}{\longmapsto} 
  (f_1,\ldots,f_N), 
\eeq
each map being defined by (\ref{eq:Atilde-B-lj-zj}) 
and (\ref{eq:fn-lj-zj-v}) respectively.  Moreover, 
whereas $\gamma_1$ a locally invertible polynomial map, 
$\gamma_2$ is birational; $\gamma_2^{-1}$ is a polynomial 
map defined by (\ref{eq:Atilde-fn}), (\ref{eq:B-fn}) 
and $v = (f_1 + \cdots + f_N)/2$.  

The Poisson structures (\ref{eq:PB-lj-zj-v}) 
and (\ref{eq:PB-fn}) induce, via these maps, 
two Poisson structures on the space in the middle. 
We show bellow that the two induced Poisson structures 
are the same.  This implies that the Poisson structures 
on the both ends are locally equivalent.

\subsubsection{Poisson algebra induced by (\ref{eq:PB-lj-zj-v})}

We first examine the Poisson structure induced by 
(\ref{eq:PB-lj-zj-v}).  Since $v$ is obviously 
a Casimir function, the problem is to calculate 
the Poisson brackets of $\atilde_j$'s and $b_j$'s.  
This calculation can be done most neatly in terms 
of generating functions, i.e., $\Atilde(\lambda)$ 
and $B(\lambda)$.  The outcome is a kind of 
``quadratic Poisson algebra'' as follows.  

\begin{prop}
The Poisson structure of (\ref{eq:PB-lj-zj-v}) 
induces a quadratic Poisson algebra of 
$\Atilde(\lambda)$ and $B(\lambda)$. 
The induced Poisson brackets read 
\beq
  \{\Atilde(\lambda),\Atilde(\mu)\} 
  = \{B(\lambda),B(\mu)\} 
  = 0, \nonumber \\
  \{\Atilde(\lambda),B(\mu)\} 
  = \frac{B(\lambda)\Atilde(\mu) - \Atilde(\lambda)B(\mu)}
         {\lambda - \mu}, 
  \label{eq:PB-Atilde-B}
\eeq
where $\lambda$ and $\mu$ are understood 
to be independent parameters.  
\end{prop}

\proof 
The proof consists of several steps.  

i) Since $\{\lambda_j,\lambda_k\} = \{\lambda_j,v\} = 0$, 
it is obvious that $\{B(\lambda), B(\mu)\} = 0$.  

ii)  Let us substitute $z_k = \Atilde(\lambda_j)\lambda_j$ 
in the fundamental relations $\{\lambda_j,z_k\} = \delta_{jk}z_k$. 
Since 
\beqnn
  \{\lambda_j,\Atilde(\lambda_k)\} 
  = \{\lambda_j,\Atilde(\lambda)\}|_{\lambda=\lambda_k} 
    + \{\lambda_j,\lambda_k\}\Atilde'(\lambda_k)  
  = \{\lambda_j,\Atilde(\lambda)\}|_{\lambda=\lambda_k}, 
\eeqnn
this yields the equalities 
\beqnn
  \{\lambda_j,\Atilde(\lambda)\}|_{\lambda=\lambda_k} 
  = \delta_{jk}\Atilde(\lambda_k) 
\eeqnn
for $k = 1,\ldots,g$.  The interpolation formula (\ref{eq:Lag}) 
can be used to assemble them to a single relation: 
\beqnn
  \{\lambda_j,\Atilde(\lambda)\} 
  = \frac{\Atilde(\lambda_j)B(\lambda)}
         {B'(\lambda_j)(\lambda - \lambda_j)}. 
\eeqnn

iii)  We now consider the obvious equality 
$\{B(\lambda_j),\Atilde(\lambda)\} = 0$.  
The left hand side can be expanded as 
\beqnn
  \{B(\lambda_j),\Atilde(\lambda)\} 
  = \{B(\mu),\Atilde(\lambda)\}|_{\mu=\lambda_j} 
    + \{\lambda_j,\Atilde(\lambda)\}B'(\lambda_j). 
\eeqnn
Therefore, applying the last formula of step ii 
to the second term on the right hand side, 
we obtain the equalities 
\beqnn
  \{B(\mu),\Atilde(\lambda)\}|_{\mu=\lambda_j} 
  = - \{\lambda_j,\Atilde(\lambda_j)\}B'(\lambda_j) 
  = - \frac{\Atilde(\lambda_j)B(\lambda)}{\lambda - \lambda_j} 
\eeqnn
for $j = 1,\ldots,g$.  Now the interpolation formula is 
invoked once again to pack them into the single relation: 
\beqnn
  \{B(\mu),\Atilde(\lambda)\} 
  = - \sum_{j=1}^g 
      \frac{\Atilde(\lambda_j)B(\lambda)B(\mu)} 
      {B'(\lambda_j)(\lambda - \lambda_j)(\mu - \lambda_j)}. 
\eeqnn
Substituting 
\beqnn
  \frac{1}{(\lambda - \lambda_j)(\mu - \lambda_j)} 
  = \frac{1}{\lambda - \mu}
    \left( \frac{1}{\mu - \lambda_j} 
         - \frac{1}{\lambda - \lambda_j} \right), 
\eeqnn
we can split the right hand side of the last equality 
into the difference of 
\beqnn
  \sum_{j=1}^g 
    \frac{\Atilde(\lambda_j)}{B'(\lambda_j)(\lambda - \lambda_j)} 
  = \frac{\Atilde(\lambda)}{B(\lambda)} 
    - \frac{\atilde_0}{b_0} 
\eeqnn
and the same quantity with $\lambda$ replaced with $\mu$.  
This yields the expression of $\{A(\lambda),B(\mu)\}$ 
in (\ref{eq:PB-Atilde-B}).  

iv)  Let us now rewrite the fundamental relations 
$\{z_j,z_k\} = 0$ in the same way.   Substituting 
$z_j = \Atilde(\lambda_j)\lambda_j$ and 
$z_k = \Atilde(\lambda_k)\lambda_k$, we encounter 
the quantity 
\beqnn
  \{\Atilde(\lambda_j),\Atilde(\lambda_k)\} 
  &=& \{\lambda_j,\Atilde(\lambda_k)\}\Atilde'(\lambda_j) 
    + \{\Atilde(\lambda),\lambda_k\}|_{\lambda=\lambda_j}\Atilde'(\lambda_k) 
  \nonumber \\
  && \mbox{}
    + \{\Atilde(\lambda),\Atilde(\mu)\}|_{\lambda=\lambda_j,\mu=\lambda_k}. 
\eeqnn
The first term on the right hand side is equal to 
$\delta_{jk}\Atilde(\lambda_k)\Atilde'(\lambda_k)$,   
which cancels with the second term, because the latter 
can be calculated as 
\beqnn
  \{\Atilde(\lambda),\lambda_k\}|_{\lambda=\lambda_j}\Atilde'(\lambda_k)
  = - \lim_{\lambda\to\lambda_j} \frac{\Atilde(\lambda_j)B(\lambda)}
         {B'(\lambda_k)(\lambda- \lambda_k)}\Atilde'(\lambda_k) 
  = - \delta_{jk}\Atilde(\lambda_k)\Atilde'(\lambda_k). 
\eeqnn
We can thus deduce that 
$\{\Atilde(\lambda),\Atilde(\mu)\}|_{\lambda=\lambda_j,\mu=\lambda_k} = 0$ 
for $j,k = 1,\ldots,g$. This implies that 
$\{\Atilde(\lambda),\Atilde(\mu)\} = 0$.  
\qed

\subsubsection{Poisson algebra induced by (\ref{eq:PB-fn})} 

We now calculate the Poisson brackets of 
$\Atilde(\lambda)$ and $B(\lambda)$ with regard 
to the Poisson structure of (\ref{eq:PB-fn}).  
The three-term recursion relations 
play a key role here again.  

\begin{prop}
The polynomials $F_n(\lambda)$, $n = 1,\ldots,N-2$, 
pairwise satisfy the Poisson relations 
\beq
  \{F_{n+1}(\lambda),F_{n+1}(\mu)\} 
  = \{F_{n+2}(\lambda),F_{n+2}(\mu)\} 
  = 0,  \nonumber \\
  \{F_{n+1}(\lambda),F_{n+2}(\mu)\} 
  = \frac{F_{n+2}(\lambda)F_{n+1}(\mu) - F_{n+1}(\lambda)F_{n+2}(\mu)}
    {\lambda - \mu} 
  \label{eq:PB-Fn} 
\eeq
with regard to the Poisson structure of (\ref{eq:PB-fn}).  
In particular, this induces the same Poisson algebra 
(\ref{eq:PB-Atilde-B}) as the one induced by the Poisson 
structure of (\ref{eq:PB-lj-zj-v}).  
\end{prop}

\proof
We prove (\ref{eq:PB-Fn}) by induction 
that starts from $n = N-3$ and proceeds downward 
to $n = 0$.  In the case where $n = N-3$, 
(\ref{eq:PB-Fn}) is a statement for 
$F_{N-2}(\lambda) = f_{N-2}f_{N-1} + \lambda + \beta_{N-2}$ 
and $F_{N-1}(\lambda) = f_{N-1}$, 
which one can directly confirm.  
Let us now assume (\ref{eq:PB-Fn}) and show that 
the same Poisson relations with $n$ replaced by 
$n - 1$ are satisfied.  More explicitly, our task 
is to derive the Poisson relations 
\beq
  \{F_n(\lambda),F_n(\mu)\} = 0, \; 
  \{F_n(\lambda),F_{n+1}(\mu)\} 
  = \frac{F_{n+1}(\lambda)F_n(\mu) - F_n(\lambda)F_{n+1}(\mu)} 
    {\lambda - \mu} 
  \label{eq:PB-Fn-next}
\eeq
of the next stage from (\ref{eq:PB-Fn}).  
This is indeed achieved by substituting 
\beqnn
  F_n(\lambda) 
  = f_nF_{n+1}(\lambda) + (\lambda + \beta_n)F_{n+2}(\lambda), \; 
  F_n(\mu) 
  = f_nF_{n+1}(\mu) + (\mu + \beta_n)F_{n+2}(\mu),
\eeqnn
and doing calculations.  Let us illustrate 
this procedure for the second part of 
(\ref{eq:PB-Fn-next}). 
Upon substitution, the left hand side of 
this Poisson relation can be expanded as 
\beqnn
  \{F_n(\lambda),F_{n+1}(\mu)\} 
  &=& \{f_n,F_{n+1}(\mu)\}F_{n+1}(\lambda) 
    + f_n \{F_{n+1}(\lambda),F_{n+1}(\mu)\} \nonumber \\
  && \mbox{} 
    + (\lambda + \beta_n)\{F_{n+2}(\lambda),F_{n+1}(\mu)\}. 
\eeqnn
The first Poisson bracket on the right hand side 
can be calculated as 
\beqnn
  \{f_n, F_{n+1}(\mu)\} 
  = - \frac{\rd F_{n+1}(\mu)}{\rd f_{n+1}} 
  = - F_{n+2}(\mu). 
\eeqnn
The other Poisson brackets can be handled 
by the Poisson relations (\ref{eq:PB-Fn}) 
that we have assumed.  One can now confirm 
the first part of (\ref{eq:PB-Fn-next}) 
by straightforward calculations.  
The first part of (\ref{eq:PB-Fn-next}) 
can be verified in much the same way.  
\qed 

We have thus arrived at the following result. 

\begin{theorem}
The rational map connecting the spectral 
Hamiltonian system and the Noumi-Yamada system 
is a Poisson map with regard to the Poisson 
structures defined by (\ref{eq:PB-lj-zj-v}) 
and (\ref{eq:PB-fn}).  
\end{theorem}

\newsection{Conclusion}

Although Adler's Lax pair for periodic dressing chains 
has an unusual structure, we have been able to apply 
the usual method for isospectral and isomonodromic 
deformations with minimal modifications.  Firstly, 
the spectral curve and the spectral Darboux coordinates 
can be constructed from the transition matrix around 
the periodic chain.  Secondly, the Lax equation of 
the dressing chain can be converted to a Hamiltonian 
system --- the spectral Hamiltonian system --- 
in the spectral Darboux coordinates.  Thirdly, 
the dynamical variables of the dressing chain can be 
reconstructed from the spectral Darboux coordinates 
by an algebraic procedure.  As a byproduct, 
we have been able to confirm that the natural 
symplectic structure of the spectral Hamiltonian system 
is consistent with one of the previously known 
Poisson structures of the periodic dressing chain 
or of the Noumi-Yamada system.  

It is significant that our ``spectral'' description 
of the Hamiltonian structure has turned out to be 
closely connected with Noumi and Yamada's Lax pair. 
This seems to indicate that one can reformulate 
everything in the language of Noumi and Yamada's 
Lax pair.  In this respect, a very intriguing problem 
is to develop a similar spectral description of 
the sixth Painlev\'e equation using Noumi and Yamada's 
new Lax pair \cite{bib:No-Ya3}. 

It should be stressed that we have been able to 
digest just a small part of the full list of 
various dressing or Darboux chains.  For instance, 
Adler's work \cite{bib:Adler2} deals with some other 
Painlev\'e equations.  Moreover, Recent work of 
Willox and Hietarinta \cite{bib:Wi-Hi} seems to 
cover a broader class of Darboux chains.

\subsection*{Acknowledgements}

This work is partially supported by 
the Grant-in-Aid for Scientific Research 
(No. 12640169 and No. 14540172) from 
the Ministry of Education, Culture, Sports 
and Technology.

\startappendix

\newsection{Interpolation formulae}

Let $B(\lambda)$ be a polynomial of the form 
\beqnn
  B(\lambda) 
  = \sum_{\ell=0}^g b_\ell\lambda^{g-\ell} 
  = b_0 \prod_{j=1}^g (\lambda - \lambda_j), 
\eeqnn
and assume that the zeroes $\lambda_j$ are mutually distinct. 
The simplest form of Lagrange's interpolation formula 
reads 
\beq
  \frac{F(\lambda)}{B(\lambda)} 
  = \sum_{j=1}^g 
    \frac{F(\lambda_j)}{B'(\lambda_j)(\lambda - \lambda_j)},  
  \label{eq:Lag}
\eeq
which holds for for any polynomial  $F(\lambda)$ of 
degree less than $g$. If the degree of $F(\lambda)$ 
is greater than or equal to $g$, extra terms appear 
on the right hand side as 
\beq
  \frac{F(\lambda)}{B(\lambda)} 
  = \sum_{j=1}^g 
    \frac{F(\lambda_j)}{B'(\lambda_j)(\lambda - \lambda_j)} 
    + G(\lambda), 
  \label{eq:Lag-mod} 
\eeq
where $G(\lambda)$ is a polynomial. 

Let us consider, in particular, the case where 
$F(\lambda) = \lambda^\ell$, $\ell = 0,\cdots,g-1$.  
This leads to the identities 
\beqnn
  \frac{\lambda^\ell}{B(\lambda)} 
  = \sum_{j=1}^g 
    \frac{\lambda_j^\ell}{B'(\lambda_j)(\lambda - \lambda_j)}. 
\eeqnn
Picking out the residue of both hand sides 
at $\lambda = \infty$ gives the formula 
\beq
  \sum_{j=1}^g \frac{\lambda_j^\ell}{B'(\lambda_j)} 
  = \frac{\delta_{\ell,g-1}}{b_0} 
  \quad (\ell = 0,\cdots,g-1), 
  \label{eq:Lag-v1}
\eeq
which has been used in the theory of 
finite-band integration.  

Another application is an inversion formula of a system 
of linear relations with Vandermonde coefficients.  
Consider the linear relations 
\beq
  \sum_{\ell=1}^g \xi_\ell \lambda_j^{g-\ell} = \eta_j 
  \quad (j = 1,\cdots,g). 
\eeq
The problem is to solve these equations for $\xi_\ell$'s. 
A clue is the auxiliary function  
$\xi(\lambda) = \sum_{\ell=1}^g \xi_\ell \lambda^{g-\ell}$.  
This is a polynomial of degree less than $g$, 
and satisfies the interpolation relations 
$\xi(\lambda_j) = \eta_j$ for $j = 1,\cdots,g$.  
The the interpolation formula thereby yields 
\beqnn
  \xi(\lambda) 
  = \sum_{j=1}^g 
      \frac{\eta_j B(\lambda)}{B'(\lambda_j)(\lambda - \lambda_j)} 
  = - \sum_{j=1}^g 
        \frac{\eta_j}{B'(\lambda_j)}
        \frac{\rd B(\lambda)}{\rd \lambda_j}. 
\eeqnn
Extracting the coefficients of the polynomials 
on both hand sides gives the inversion formula 
\beq
  \xi_\ell 
  = - \sum_{j=1}^g 
      \frac{\eta_j}{B'(\lambda_j)}
      \frac{\rd b_\ell}{\rd \lambda_j} 
  \quad (\ell = 1,\cdots,g). 
  \label{eq:Lag-v2}
\eeq

\newsection{Equivalence of even-periodic dressing chain 
and Noumi-Yamada system} 

The identity (\ref{eq:dot-vn-identity}), 
combined with (\ref{eq:dchain}), yields 
the quadratic constraint 
\beq
  \sum_{n=1}^{2g+2}(-1)^n(v_{n+1}^2 - v_n^2 + \alpha_n) = 0  
  \label{eq:vn-quad-constr}
\eeq
on $v_n$'s.  This constraint can be rewritten as 
\beqnn
  \sum_{k=1}^{g+1}(v_{2k}^2 - v_{2k-1}^2) 
  = \frac{1}{2}\sum_{n=1}^{2g+2}(-1)^n\alpha_n. 
\eeqnn
Substituting $v_{2k}^2 - v_{2k-1}^2 
= (v_{2k} - v_{2k-1})f_{2k-1}$, the quadratic 
constraint turns into the linear constraint 
\beq
  \sum_{k=1}^{g+1} (v_{2k} - v_{2k-1})f_{2k-1} 
  = \frac{1}{2}\sum_{n=1}^{2g+2}(-1)^n\alpha_n. 
  \label{eq:vn-lin-constr}
\eeq

One can use this linear constraint to remove 
the ambiguity for solving (\ref{eq:fn-vn}) 
as follows.  Firstly, solve (\ref{eq:fn-vn}) 
for $v_2,v_3,\ldots$, successively as 
\beq
   v_2 = f_1 - v_1, \; 
   v_3 = f_2 - v_2 = f_2 - f_1 + v_1, \; \ldots, 
   \nonumber \\
   v_n = f_{n+1} - f_n + \cdots + (-1)^nf_1 - (-1)^nv_1. 
   \label{eq:vn-fn-v1}
\eeq
One can thereby eliminate $v_n$'s from 
(\ref{eq:vn-lin-constr}) and obtain the equality 
\beqnn
  \left(\sum_{k=1}^{g+1}f_{2k-1}\right)^2 
  - 2 \sum_{1\le j\le k\le g}f_{2j}f_{2k+1} 
  - 2 \sum_{k=1}^{g+1}v_1f_{2k-1}
  = \frac{1}{2}\sum_{n=1}^{2g+2}(-1)^n\alpha_n. 
\eeqnn
Solving for $v_1$ gives the expression 
\beq
  v_1 
  = \frac{1}{2v}\left(v^2 
      - 2\sum_{1\le j\le k\le g}f_{2j}f_{2k+1} 
      - \frac{1}{2}\sum_{n=1}^{2g+2}(-1)^n\alpha_n\right) 
\eeq
of $v_1$ as a rational function of $f_n$'s and $v$.  
This, in turn, determines the other $v_n$'s by 
(\ref{eq:vn-fn-v1}).  

Thus, although the linear relation (\ref{eq:fn-vn}) 
itself cannot be solved for $v_n$'s uniquely, 
the quadratic constraint (\ref{eq:vn-quad-constr}) 
removes the ambiguity and yields a rational expression 
of $v_1,\ldots,v_{2g+2}$ in terms of $f_n$'s and $v$.  
In other words, the $(2g + 2)$-periodic dressing chain 
is connected with the Noumi-Yamada system of 
$A^{(1)}_{2g+1}$ type by a birational change of coordinates.

\newpage
%%%%%%%%%%%%%%%%%%%%%%%%%%%%%%%%%%%%%%%%%%%%%%%%%%%%%%

\end{document}